\documentclass[12pt,a4paper,onecolumn, numberedappendix]{emulateapj}
\pdfoutput=1
\usepackage{natbib}
\shorttitle{Machine-Learned Classification of Variable Stars}
\shortauthors{}
\usepackage{epsfig,graphicx,color}

\begin{document}

\title{On Machine-Learned Classification of Variable Stars with Sparse and Noisy Time-Series Data}
\author{
Joseph W. Richards\altaffilmark{1,2,*},
Dan L. Starr\altaffilmark{1},
Nathaniel R. Butler\altaffilmark{1},
Joshua S. Bloom\altaffilmark{1},
John M. Brewer\altaffilmark{3},
Arien Crellin-Quick\altaffilmark{1},
Justin Higgins\altaffilmark{1},
Rachel Kennedy\altaffilmark{1},
Maxime Rischard\altaffilmark{1}
}
\altaffiltext{1}{Astronomy Department, University of California, Berkeley, CA, 94720-7450, USA}
\altaffiltext{2}{Statistics Department, University of California, Berkeley, CA, 94720-7450, USA}
\altaffiltext{3}{Astronomy Department, Yale University, New Haven, CT 06520-8101, USA}
\altaffiltext{*}{E-mail: {\tt jwrichar@stat.berkeley.edu}}

\begin{abstract}
With the coming data deluge from synoptic surveys, there is a growing
need for frameworks that can quickly and automatically produce
calibrated classification probabilities for newly-observed variables
based on a small number of time-series measurements.  In this paper,
we introduce a methodology for variable-star classification, drawing
from modern machine-learning techniques.  We describe how to
homogenize the information gleaned from light curves by selection and
computation of real-numbered metrics (ÒfeaturesÓ), detail methods to
robustly estimate periodic light-curve features, introduce
tree-ensemble methods for accurate variable star classification, and
show how to rigorously evaluate the classification results using cross
validation.  On a 25-class data set of 1542 well-studied variable stars, we
achieve a 22.8\% overall classification error using the random forest
classifier; this represents a 24\% improvement over the best previous
classifier on these data.  This methodology is effective for
identifying samples of specific science classes:  for pulsational
variables used in Milky Way tomography we obtain a discovery
efficiency of 98.2\% and for eclipsing systems we find an efficiency
of 99.1\%, both at 95\% purity.  We show that the random forest (RF)
classifier is superior to other machine-learned methods in terms of
accuracy, speed, and relative immunity to features with no useful
class information; the RF classifier can also be used to estimate
the importance of each feature in classification.  Additionally, we
present the first astronomical use of hierarchical classification
methods to incorporate a known class taxonomy in the classifier, which
further reduces the catastrophic error rate to 7.8\%.  Excluding
low-amplitude sources, our overall error rate improves to 14\%, with a
catastrophic error rate of 3.5\%.
\end{abstract}

\keywords{stars: variables: general -- methods: data analysis -- methods: statistical -- techniques: photometric}

\section{Introduction}
\label{sec:intro}

Variable star science (e.g., \citealt{em08}) remains at the core of many of the central pursuits in astrophysics: {\it pulsational} sources probe stellar structure and stellar evolution theory, {\it eruptive and episodic} systems inform our understanding of accretion, stellar birth, and mass loss, and {\it eclipsing} systems constrain mass transfer, binary evolution, exoplanet demographics, and the mass-radius-temperature relation of stars. Some eclipsing systems and many of the most common pulsational systems (e.g., RR Lyrae, Cepheids, and Mira variables) are the fundamental means to determine precise distances to clusters, to relic streams of disrupted satellites around the Milky Way, and to the local group of galaxies. They anchor the measurement of the size scale of the Universe. See \citet{2009astro2010S.307W} for a recent review.

The promise of modern synoptic surveys \citep{2007AJ....134..973I}, such as the Large Synoptic Survey Telescope (LSST), is the promise of discovery of many new instances of variable stars \citep{2007AJ....134.2236S}, some to be later studied individually with greater photometric and spectroscopic scrutiny\footnote{High-precision photometry missions (Kepler, MOST, CoRoT, etc.) are already challenging the theoretical understanding of the origin of variability and the connection of some specific sources to established classes of variables.} and some to be used as ensemble probes to larger volumes. New classes (with variability reflecting physics not previously seen) and rare instances of existing classes of variables are almost certainly on the horizon (e.g., \citealt{2007AJ....134.2398C}).

Classification of variable stars---the identification of a certain variable with a previously identified group (``class'') of sources presumably of the same physical origin---presents several challenges. First, time-series data alone (i.e., without spectroscopy) provides  an incomplete picture of a given source: this picture is even less clear the more poorly sampled the light curve is both in time and in precision. Second, on conceptual grounds, the observation of variability does not directly reveal the underlying physical mechanisms responsible for the variability. What the totality of the characteristics {\it are} that define the nature of the variability may in principle be known at the statistical level. But {\it why} that variability is manifest relies on an imperfect mapping of an inherently incomplete physical model to the data. (For example, the periodic dimming of a light curve may be captured with a small number of observable parameters but the inference that that source is an eclipsing one requires a theoretical framework.) This intermingling of observation and theory has given rise to a taxonomy of variable stars (for instance, defined in the GCVS\footnote{General Catalog of Variable Stars, {\tt http://www.sai.msu.su/groups/cluster/gcvs/gcvs/}}) that is based on an admixture of phenomenology and physics. Last, on logistical grounds, the data volume of time-series surveys may be too large for human-intensive analysis, follow-up, and classification (which benefits from domain-specific knowledge and insight).

While the data deluge problem suggests an obvious role for computers in classification\footnote{Not discussed herein are the challenges associated with {\it discovery} of variability. See \citet{2009MNRAS.400.1897S} for a review.}, the other challenges also naturally lend themselves to algorithmic and computational solutions. Individual light curves can be automatically analyzed with a variety of statistical tools and the outcome of those analyses can  be handled with machine-learning algorithms that work with existing taxonomies (however fuzzy the boundary between classes) to produce statistical statements about the source classification. Ultimately, with a finite amount of time-series data we wish to have well-calibrated probabilistic statements about the physical origin and phenomenological class of that source.

While straightforward in principle, providing a machine-learned classifier that is accurate, fast, and well-calibrated is an extraordinarily difficult task on many fronts (see discussion in \citealt{2008AIPC.1082..257E}). There may be only a few instances of light curves in a given class (``labelled data'') making training and validation difficult. Even with many labelled instances, in the face of noisy, sometimes spurious, and sparsely sampled data, there is a limit to the statistical inferences that can be gleaned from a single light curve. Some metrics (called, in machine-learning parlance,``features'') on the light curve may be very sensitive to the signal-to-noise of the data and others, particularly frequency-domain features, may be sensitive to the precise cadences of the survey (\S \ref{ss:surveydependence}). For computationally intensive feature generation (e.g., period searches) fast algorithms may be preferred over slower but more robust algorithms.  

Machine learning in variable star classification has been applied to several large time-series datasets \citep{2004AJ....128.2965W,2007arXiv0712.2898W,2007debo,2008AN....329..288M,2009A&A...494..739S,2010blom}. A common thread for most previous work is application of a certain machine-learning framework to a single survey. And, most often, the classification is used to distinguish/identify a small set of classes of variables (e.g., Miras and other red giant variability). \citet{2007debo} was the first work to tackle the many-class ($>20$) problem with multiple survey streams. \citet{2007debo} also explored several classification frameworks and quantitatively compared the results.

The purpose of this work is to build a many-class classification framework by exploring in detail each aspect of the classification of variable stars: proper feature creation and selection in the presence of noise and spurious data (\S \ref{sec:features}), fast and accurate classification (\S \ref{sec:methods}), and improving classification by making use of the taxonomy. We present a  formalism for evaluating the results of the classification in the context of expected statistical risk for classifying new data. We use data  analyzed by Debosscher et al.\ to allow us to make direct comparison with those results (\S \ref{sec:results}). Overall, we find a 24\% improvement in the misclassification rate with the same data. The present work only makes use of metrics derivable from time-domain observations in a single bandpass; color information and context (i.e., the location of the variable in the Galaxy and with respect to other catalog sources) are not used. In future work, we will explore how machine-learned classifiers can be applied across surveys (with different characteristics) and how context and time-domain features can be used in tandem to improve overall classification. \\

\pagebreak

\section{Homogenizing Light Curves: Feature Generation}
\label{sec:features}

Classification fundamentally relies upon the ability to recognize and quantify the differences between light curves. To build a classifier, many {\it instances} of light curves are required for each class of interest. These labelled instances are used in the {\it training} and {\it testing} process (\S \ref{sec:methods}). Since the data are not, in general, sampled at regular intervals nor are all instances of a certain class observed with the same number of epochs and signal-to-noise, identifying the differences directly from the time-series data is exceedingly challenging both conceptually and computationally (cf.\ \citealt{2004SPIE.5200...79E}). Instead we homogenize the data by transforming each light curve into a set of real-number line features using statistical and model-specific fitting procedures. For variable stars, features fall into two broad categories: those related to the period of the source (and harmonics) and those that are not. Which features to use (and not use) is an important question that we will address herein. We also address the effects of (implicit) correlation of certain features in affecting the classification model.

Appendix \ref{app:features} provides an account of the non-periodic features used in this present work; many of these are simple statistics on the distribution of the fluxes (e.g., median absolute deviation and min-max amplitude) and some are domain specific (such as a feature that captures how much a source varies like the damped random walk seen in quasars; \citealt{2010butl}). Since the period and periodic signatures of a variable are such crucial quantitative measurements, yet tend to be difficult to infer from simple prescriptions, we review the algorithms we employ to compute these features.

\subsection{Robust estimation of periodic features}
\label{sec:perfeat}

\subsubsection{A Fast Period Search Including Measurement Uncertainty}

We model the photometric magnitudes of variable stars 
versus time $t$ as a superposition of sines and cosines, starting
from the most basic form:
\begin{equation}
y_i(t|f_i) = a_i \sin(2\pi f_i t) + b_i \cos(2\pi f_i t) + b_{i,\circ},
\label{eq:basic}
\end{equation}
where $a$ and $b$ are normalization constants for the sinuoids
of frequency $f_i$, and $b_{i,\circ}$ is the magnitude offset.
For each variable star, we record at each epoch, $t_k$, a photometric 
magnitude, $d_k$, and its uncertainty, $\sigma_k$.   To search 
for periodic variations in these data, we fit
 (\ref{eq:basic}) by minimizing the
sum of squares
\begin{equation}
\chi^2 = \sum_k [d_k-y_i(t_k)]^2/\sigma_k^2,
\end{equation}
where $\sigma_k$ is the measurement uncertainty in data point $d_k$.
As discussed in \citet{zech09}, this least-squares fitting of
sinusoids with a floating mean and over a range of test frequencies
is closely similar to an evaluation of the well-known 
Lomb-Scargle \citep{1976lomb,1963barn,1982scar}
periodogram.  Allowing the mean to float leads to more robust
period estimates in the cases where the periodic phase is not
uniformly sampled; in these cases, the model light curve has a non-zero
mean.  (This is particularly important for searching for periods 
on timescales similar to the data span $T_{\rm tot}$.)
If we define: 
\begin{equation}
  \chi^2_{\circ} = \sum_k [d_k-\mu]^2/\sigma_k^2,
\end{equation}
with the weighted mean 
$\mu = \sum_k [d_k/\sigma_k^2] / \sum_k 1/\sigma_k^2$,
then our generalized Lomb-Scargle periodogram $P_f$ is
\begin{equation}
 P_f(f) = {(N-1) \over 2} {\chi^2_{\circ} - \chi^2_m(f) \over
\chi^2_{\circ}}, \label{eq:gls}
\end{equation}
where $\chi^2_m(f)$ is $\chi^2$ minimized with respect to
$a$, $b$, and $b_{\circ}$.
For the NULL hypothesis of
no periodic variation and a white noise spectrum, we expect $P_f$
to be $F$-distributed with 2 numerator
and $N-1$ denominator degrees of freedom.
A similar periodogram statistic and NULL
distribution is derived in \citet{gregory05} by
marginalizing over an unknown scale
error in the estimation of the uncertainties.  In the limit
of many data, the NULL distribution takes the well-known exponential
form \citep[e.g.,][]{zech09}.
For all $\sigma_i=1$,
(\ref{eq:gls}) becomes the standard Lomb-Scargle periodogram.
In addition to the benefits of allowing a floating
mean, the generalized Lomb-Scargle periodogram (\ref{eq:gls})
 has two principal advantages over the standard formula:
(1) uncertainties on the measurements are included, and because
$P_f(f)$ is a constructed from a ratio of two sums-of-squares, (2)
there can be scale errors on the determination of these uncertainties
\citep[c.f.,][]{gregory05}.

We undertake the search for periodicity in each source by 
evaluating (\ref{eq:gls}) on a linear test grid in frequency
from a minimum value of $1/T_{\rm tot}$ to a maximum value
of 20 cycles per day, in steps of $\delta f = 0.1/T_{\rm tot}$. This 
follows
closely the prescription in \citet{2007debo}, with the important
exception that we search for periods up to 20 cycles per day in
all sources, whereas \citet{2007debo} search up to a
``pseudo'' Nyquist frequency ($f_N=0.5\langle 1/\Delta T\rangle$,
where $\Delta T$ is the difference in time between observations
and $\langle \cdot \rangle$ is an average) for most sources but
allow the maximum frequency value to increase for certain
source classes.  To avoid favoring spurious high-frequency peaks in the periodogram, we subtract a mild penalty
of $\log{f/f_N}$ above $f_N$ from $P_f(f)$ above $f=f_N$.
Significance of the highest peak is the evaluated from $P_f(f)$.
We apply an approximate correction for the number of search trials
using the prescription of \citet{hb86}, although we note that
numerical simulations suggest these significance estimates 
underestimate the number of true trials and are
uniformly high by $1$--$2\sigma$.

Standard ``fast'' implementations of the Lomb-Scargle periodogram
\citep[e.g.,][]{nr}, which scale with the number of frequency bins
$N_f$ as $N_f\log{N_f}$, are not particularly fast for our purposes.
This is because we wish to sample relatively few data ($N\sim 100$) on
a very dense, logarithmic frequency grid $N_f\sim 10^6$.  It is
more fruitful to pursue algorithms which scale with $N$.  We find that
standard implementations are sped up by a factor of $\sim 10$
by simply taking care in calculating the sines and cosines that are
necessary to tabulate (\ref{eq:gls}).  Instead of calculating
all the sines and cosines at a given time point at each of the $N_f$ frequency
bins, we calculate sine and cosine only once at $f=\delta f$ and
then use trigonometric identities (i.e., successive rotations by 
an angle $2 \pi\delta f t_i$) to determine the sines and cosines at 
higher frequencies.

\subsubsection{Fitting Multiple Periods}

Following \citet{2007debo}, we fit each light curve with a linear
term plus a harmonic sum of sinusoids:
\begin{equation}
y(t) = ct + \sum_{i=1}^3 \sum_{j=1}^4 y_i(t|jf_i),
\label{eq:full}
\end{equation}
where each of 3 test frequencies $f_i$ are allowed to have 4 harmonics
at frequencies $f_{i,j} = j f_i$.
The 3 test frequencies $f_i$ are found iteratively, by successively
finding the periodogram peaks in $P_f(f)$. This procedure assumes
a model with no harmonics, only the fundamental.  We then relax 
this assumption
by fitting for the fundamental plus 3 harmonics of the fundamental.
Unlike \citet{2007debo} who
fit for the linear term prior to fitting sines and cosines,
we evaluate the best-fit linear term at each
test frequency during
the search for the first periodogram peak, using a modification of
(\ref{eq:gls}) to fit for the linear trend coefficient $c$
at the same time as the constants $a_0$, $b_0$, and $b_{0,\circ}$.
The value for $c$ is updated when we then calculate the 4-harmonic
fit (plus linear term) around the first best-fit frequency.
The quantity $\chi^2_{\circ}$ is updated after the $i^{\rm th}$ iteration
and subtraction
of each 4-harmonic component is performed prior to calculating $P_f(f)$.

In reporting the values from the fit of (\ref{eq:full}),
we ignore the constant offsets $b_{i,\circ}$.  We translate the
sinusoid coefficients into an amplitude and a phase:
\begin{eqnarray}
 A_{i,j} &=& \sqrt{ a_{i,j}^2+b_{i,j}^2 } \\
 PH_{i,j} &=& \tan^{-1}( b_{i,j},a_{i,j} ).
\end{eqnarray}
Here $A_{i,j}$ ($PH_{i,j}$) is the amplitude (phase) of
the $j^{\rm th}$ harmonic of the $i^{\rm th}$ frequency component.
Following \citet{2007debo},
we correct the phases $PH_{i,j}$ to relative phases with respect
to the phase of the first component $PH'_{i,j} = PH_{i,j}-PH_{00}$.
This is to preserve comparative utility in the phases for
multiple sources by dropping a non-informative phase offset for each
source.
All phases are then remapped to the interval $|-\pi,+\pi|$.

A list and summary of all of the period features used in our analysis is found in table \ref{tab:LS} in Appendix \ref{app:features}.

\begin{figure}
\begin{center}
$\begin{array}{cc}
\multicolumn{1}{l}{\mbox{\bf (A)}} & \multicolumn{1}{l}{\mbox{\bf (B)}} \\ 	
\includegraphics[angle=0,width=3.0in]{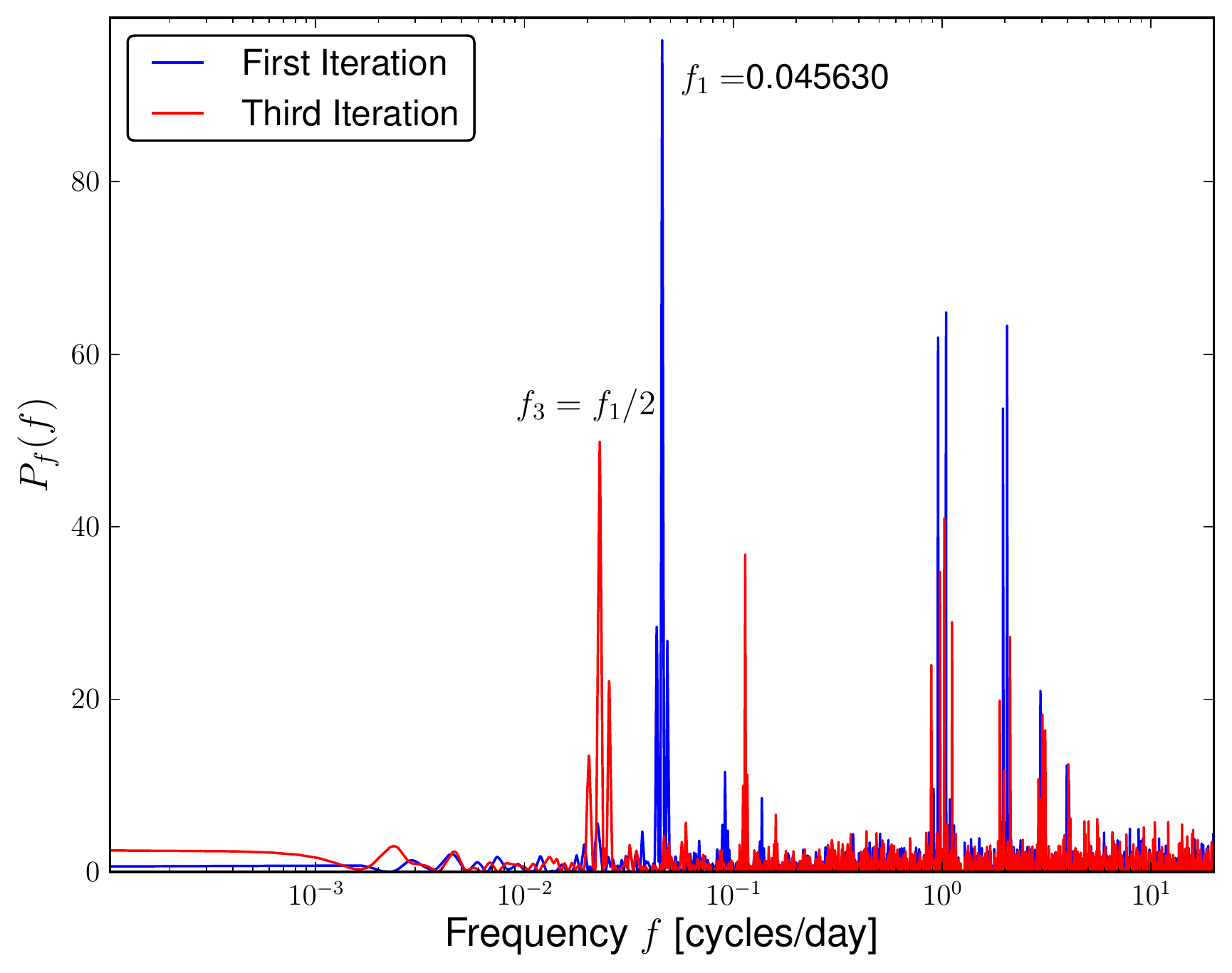} &
\includegraphics[angle=0,width=3.0in]{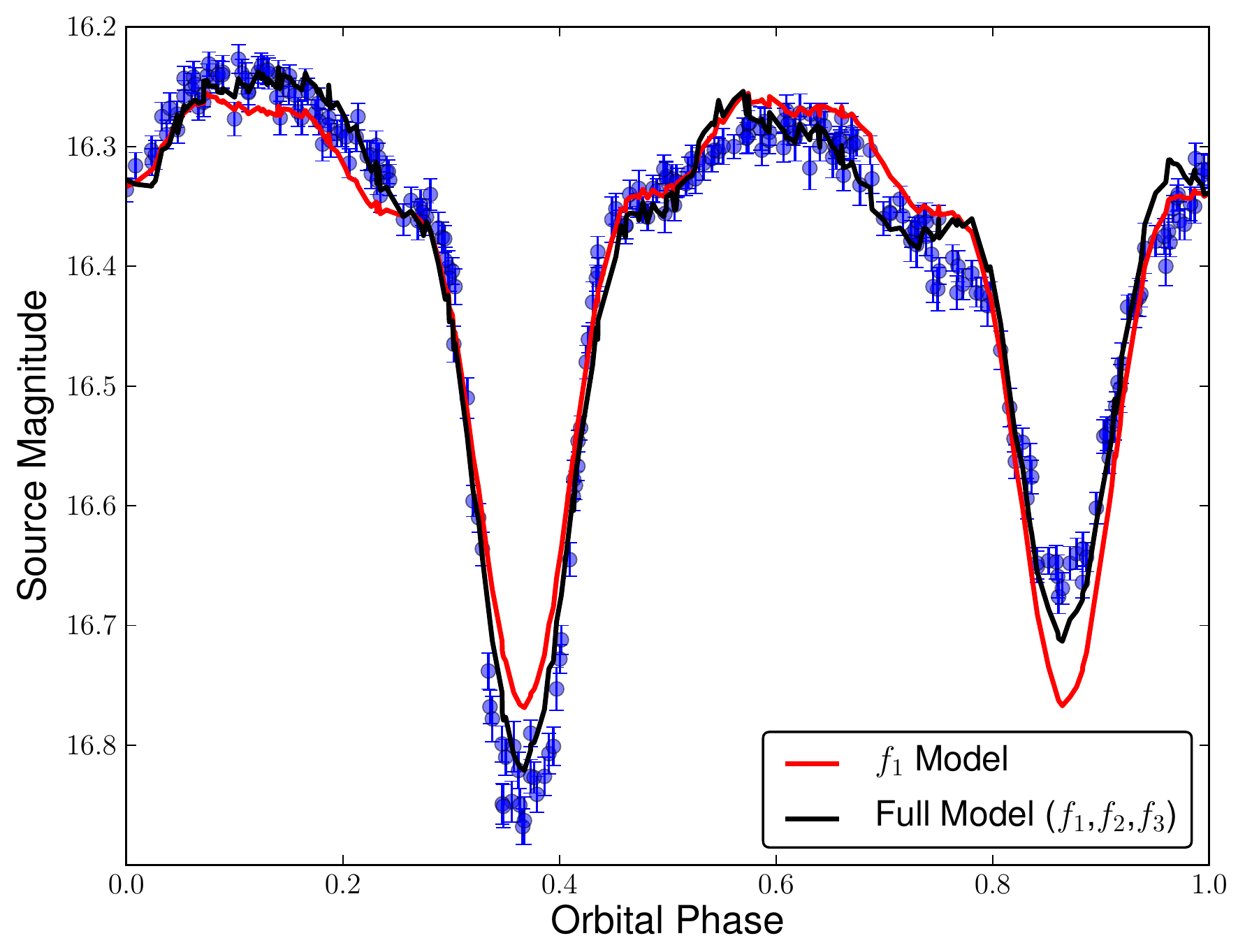}\\
\end{array}$
\end{center}
\label{figure:folded}
\caption{(A) Generalized Lomb-Scargle Periodogram $P_f(f)$ for
an eclipsing source in the sample.  Plotted in blue is the first
iteration to find peak frequency $f_1$, which is twice the orbital frequency
$f_3$ (third iteration plotted in red).  In this case, the second iteration
yielded $f_2=3f_3$.  For eclipsing sources, our $P_f(f)$  analysis 
which utilizes a sine and cosine fit without harmonics,
tends to place the orbital period in either $f_2$ or $f_3$.
(B) The light curve folded at the orbital period $f_3$.  Over plotted
is the best-fit model considering only $f_1$ (plus 3 harmonics; in red),
which fails to account for the difference in primary and secondary eclipse
depths.  Addition of the 2nd and 3rd frequency component models (black curve)
account for the full light curve structure well.}
\end{figure}

\subsection{Non-periodic light curve features}
\label{sec:nonperfeat}

In seeking to classify variable star light curves, it may not always
be possible to characterize flux variation purely by detecting
and characterizing periodicity. We find that simple summary statistics
of the flux measurements (e.g., standard deviation, skewness, etc.)
--- determined without sorting the data in time or period phase ---
give a great deal of predictive power.  For instance, skewness is very
effective for separating eclipsing from non-eclipsing sources.  We
define (in Appendix \ref{app:features}) 20 non-periodic features and explore in \S \ref{sec:results}
their utility for source classification.  A summary of all of the non-period features used in our analysis is found in table \ref{tab:nonLS} in Appendix \ref{app:features}.

When only a small number of epochs ($\lesssim 12$) are sampled, we find
that period detection becomes unreliable: we can only rely on
crude summary statistics and contextual information to characterize these sources.  In addition,
some source classes yield non-period or multiply-periodic light curves,
wherby the non-periodic features are expected in these cases to carry
useful additional information not already contained in the periodic
features.  As an example, we apply metrics used in the time-domain
study of quasars \citep{2010butl} to aid in disentangling the
light curves of some complexly varying, long-period variables (e.g.,
semi-regular pulsating variables).\\

\section{Classification frameworks for variable stars}
\label{sec:methods}

The features extracted from a light curve give a characterization of the observed astronomical source.  We need a rigorous way of turning this information into a probabilistic statement about the science class of that source.  This is the goal of classification: given a set of sources whose science class is known, learn a model that describes each source's class probabilities as a function of its features. This model is then used to automatically predict the class probabilities, and the most likely science class, of each new source.  

Several authors have used machine-learning methods to classify variable stars using their light curves: \citet{2004bret} use Kohonen self-organizing maps, \citet{2005eyer1} use the Bayesian mixture-model classifier {\tt Autoclass} (\citealt{1996chee}) and \citet{2007debo} experiment with several methods, including Gaussian mixture models, Bayesian networks, Bayesian averaging of artificial neural networks, and support vector machines.  All of these methods have certainly enjoyed widespread use in the literature, and are a reasonable first set of tools to use in classifying variable stars.  Our major contribution in this section is to introduce tree-based classification methods---including classification and regression trees (CART), random forests, and boosted trees---for the classification of variable stars.  Tree-based classifiers are powerful because they are able to capture complicated interaction structure within the feature space, are robust to outliers, naturally handle multi-class problems, are immune to irrelevant features, easily cope with missing feature values, and are computationally efficient and scalable for large problems.  Furthermore, they are simple to interpret and explain and generally yield accurate results.  In \S \ref{sec:results} we show the superior performance of tree-based methods over the methods used in \citet{2007debo} for classifying variable stars. 

Below, we describe several tree-based classification approaches from the statistics and machine learning literature, showing how to train each classifier and how to predict science-class probabilities for each observed source.  We also introduce a suite of pair-wise voting classifiers, where the multi-class problem of variable-star classification is simplified into a set of two-class problems and the results are aggregated to estimate class probabilities.  Additionally, we outline a procedure for incorporating the known variable star class taxonomy into our classifier.  Finally, we describe a rigorous risk-based framework to choose the optimal tuning parameter(s) for each classifier, and show how to objectively assess the expected performance of each classifier through cross-validation.

\subsection{Tree-based classifiers}

Decision tree learning has been a popular method for classification and regression in statistics and machine learning for more than 20 years (\citealt{1984brei} popularized this approach).  Recently, the astronomical community has begun to use tree-based techniques, with impressive results.  For example, tree-based classifiers have been used by \citet{2005such} for SDSS object classification, by \citet{2006ball} and \citet{2009okee} for star-galaxy separation, by \citet{2007bail} to identify supernova candidates, and by several groups for supernova classification in the recent DES Supernova Photometric Classification Challenge (\citealt{2010kess}).

Tree-based learning algorithms use recursive binary partitioning to split the feature space, $\mathcal{X}$, into disjoint regions, $R_1,R_2,...,R_M$.   Within each region, the response is modeled as a constant.  Every split is performed with respect to one feature, producing a partitioning of $\mathcal{X}$ into a set of disjoint rectangles (nodes in the tree).   At each step, the algorithm selects both the feature and split point that produces the smallest impurity in the two resultant nodes.  The splitting process is repeated, recursively, on all regions to build a tree with multiple levels.

In this section we give an overview of three tree-based methods for classification: classification trees, random forest, and boosting.  We focus on the basic concepts and a few particular challenges in using these classifiers for variable-star classification.  For further details about these methods, we refer the interested reader to \citet{2009hast}.

\subsubsection{Classification trees}

To build a classification tree, begin with a training set of (feature, class) pairs $(\mathbf{X}_1,Y_1),...,(\mathbf{X}_n,Y_n)$, where $Y_i$ can take any value in $\{1,...,C\}$. At node $m$ of the tree, which represents a region $R_m$ of the feature space $\mathcal{X}$, the probability that a source with features in $R_m$ belongs to class $c$ is estimated by 
\begin{equation}
\label{eqn:cart}
\widehat{p}_{mc} = \frac{1}{N_m} \sum_{\mathbf{X}_i \in R_m} I(Y_i = c),
\end{equation}
which is the proportion of the $N_m$ training set objects in node $m$ whose science class is $c$, where $I(Y_i=c)$ is the indicator function defined to be 1 if $Y_i=c$ and 0 else.  In the tree-building process, each subsequent split is chosen amongst all possible features and split points to minimize a measure of the resultant node impurity, such as the Gini index ($\sum_{c\ne c'}^C\widehat{p}_{mc}\widehat{p}_{mc'}$) or entropy ($-\sum_{c=1}^C \widehat{p}_{mc}\log_2\widehat{p}_{mc}$).  The Gini index is the measure of choice for CART (\citealt{1984brei}), while entropy is used by the popular algorithm C4.5 (\citealt{1996quin}).  This splitting process is repeated recursively until some pre-defined stopping criterion (such as minimum number of observations in a terminal node, or relative improvement in the objective function) is reached.

Once we have trained a classification tree on the examples $(\mathbf{X}_1,Y_1),...,(\mathbf{X}_n,Y_n)$, it is simple to ingest features from a new instance, $\mathbf{X}_{\rm new}$, and predict its science class.  Specifically, we first identify which tree partition $\mathbf{X}_{\rm new}$ resides in, and then assign it a class based on that node's estimated probabilities from (\ref{eqn:cart}).   For example, if $\mathbf{X}_{\rm new} \in R_m$, then the classification-tree probability that the source is in class $c$ is
\begin{equation}
\label{eqn:cartpred}
\widehat{p}_{c}(\mathbf{X}_{\rm new}) = \widehat{p}_{mc}
\end{equation}
where $\widehat{p}_{mc}$ is defined in (\ref{eqn:cart}).  Using (\ref{eqn:cartpred}), the predicted science class is $\widehat{p}(\mathbf{X}_{\rm new}) = \arg\max_c \widehat{p}_{c}(\mathbf{X}_{\rm new})$.  Note that we are free to describe the classification output for each new source either as a vector of class probabilities or as its predicted science class.

There remains the question of how large of a tree should be grown.  A very large tree will fit the training data well, but will not generalize well to new data.  A very small tree will likely not be large enough to capture the complexity of the data-generating process.  The appropriate size of a tree ultimately depends on the complexity of model necessary for the particular application at hand, and hence should be determined by the data.  The standard approach to this problem is to build a large tree, $T$, with $M$ terminal nodes and then to \emph{prune} this tree to find the subtree $T^*$ of $T$ that minimizes a cross-validation estimate of its statistical risk (see \S \ref{ss:cv}).

\subsubsection{Random forest}

Classification trees are simple, yet powerful, non-parametric classifiers.  They work well even when the true model relating the feature space to the class labeling is complicated, and generally yield estimates with very small bias.  However, tree models tend to have high variance.  Small changes in the training set features can produce very different estimated tree structure.  This is a by-product of the hierarchical nature of the tree model: small differences in the top few nodes of a tree can produce wildly different structure as those perturbations are propagated down the tree.  To reduce the variance of tree estimates, \emph{bagging} (bootstrap aggregation, \citealt{1996brei}) was proposed to average the predictions of $B$ trees fitted to bootstrapped samples of the training data.  \emph{Random forest} (\citealt{2001brei}) is an improvement to bagging that attempts to de-correlate the $B$ trees by selecting a random subset of $m_{\rm try}$ of the input features as candidates for splitting at each node during the tree-building process.  The net result is that the final, averaged random forest model has lower variance than the bagging model, while maintaining a small bias (see \citealt{2009hast}, ch.15 for a discussion). 

To obtain a random forest classification model, we grow $B$ de-correlated classification trees.  For a new variable star, the class probabilities are estimated as the proportion of the $B$ trees that predict each class.  As in classification trees, we are free to describe each source as a vector of class probabilities or a best-guess class.  This prescription generally works well because by averaging the predictions over many bootstrapped trees, the estimated probabilities are more robust to chance variations in the original training set, and are almost always more accurate than the output of a single tree.    Another advantage to random forest is the relative robustness of the estimates to choices of the tuning parameters ($B$, $m_{\rm try}$, and the size of each tree) compared to other non-parametric classification techniques.  In practice, we use the parameter values that give minimal cross-validation risk.

\subsubsection{Boosted trees}

Boosting is a method of aggregating simple rules to create a predictive model whose performance is `boosted' over that of any of its ensemble members (\citealt{1996freu}).  In classification boosting, a sequence of simple classifiers (referred to as weak-learners) is applied, whereby in each iteration the training observations are re-weighted so that those sources which are repeatedly misclassified are given greater influence in the subsequent classifiers.  Therefore, as the iterations proceed, the classifier pays more attention to data points that are difficult to classify, yielding improved overall performance over that of each weak-learner.  The predicted class probabilities are obtained from a weighted estimate of the individual classifiers, with weights proportional to the accuracy of each classifier.

Classification trees are natural base learners in a boosting algorithm because of their simplicity, interpretability, and ability to deal with data containing outliers and missing values.  Moreover, there are efficient algorithms that can quickly estimate boosted trees using gradient boosting (\citealt{2001frie}).   It is usually sufficient to use single-split trees (so-called decision stumps) as base-learners, though in situations with more complicated interactions, bigger trees are necessary.  We use the training data to choose this tuning parameter through cross-validation.

\subsection{Measuring feature importance}
\label{ss:featimpmethod}

An additional advantage to tree-based classifiers is that, because the trees are constructed by splitting one feature at a time, they allow us to estimate the importance of each feature in the model.  A feature's importance can be deduced by, for instance, counting how often that feature is split or looking at the resultant decrease in node impurity for splits on that feature. Additionally, random forest provides a measure of the predictive strength of each feature, referred to as the \emph{variable importance}, which tells us roughly what the decrease in overall classification accuracy would be if a feature were replaced by a random permutation of its values.  Random forest has a rapid procedure for estimating variable importance via its \emph{out-of-bag} samples for each tree, those data that were not included in the bootstrapped sample.

Analyzing the feature importance is a critical step in building an accurate classification model.  By determining which features are important for distinguishing certain classes, we gain valuable insight into the physical differences between particular science classes.  Moreover, we can visualize which types of features are more predictive than others, which can inform the use of novel features or the elimination of useless features in a second-generation classifier.

\subsection{Pairwise classifiers}
\label{ss:pairwise}
A common approach for multi-class classification is to reduce the $C$-class problem into a set of $C(C-1)/2$ pairwise comparisons.  This is a viable approach because two-class problems are usually easier to solve since the class boundaries tend to be relatively simple.  Moreover, some classification methods, such as support vector machines (SVM, \citealt{2000vapn}), are designed to work only on two-class problems.  In pairwise classification, classifiers for all $C(C-1)/2$ two-class problems are constructed.  The challenge, then, is to map the output from the set of pairwise classifiers for each source (pairwise probabilities or class indicators) to a vector of $C$ class probabilities that accurately reflects the science class of that source.  This problem is referred to as pairwise coupling.

The simplest method of pairwise coupling is voting (\citealt{1990kner,1996frie}), where the class of each object is determined as the winner in a pairwise head-to-head vote.  Pairwise voting is suboptimal because it ignores the pairwise class probabilities and tends to estimate inaccurate $C$-class probabilities.  In situations where pairwise class probability estimates are available, voting is outperformed by other methods such as the Kullback-Leibler based technique of \citet{1998hast} and the approaches of \citet{2004wu}, which reduce the problem to solving a linear system of equations.  In this paper, we explore the use of both tree-based classifiers and SVMs in pairwise classifiers.  To obtain $C$-class probabilities, we employ the second pairwise coupling method introduced by \citet{2004wu}.  We refer the interested reader to that paper for a more detailed description of the method and a review of similar techniques in the literature.

\subsection{Hierarchical classification}

\begin{figure}
\begin{center}
\includegraphics[angle=0,width=5.0in]{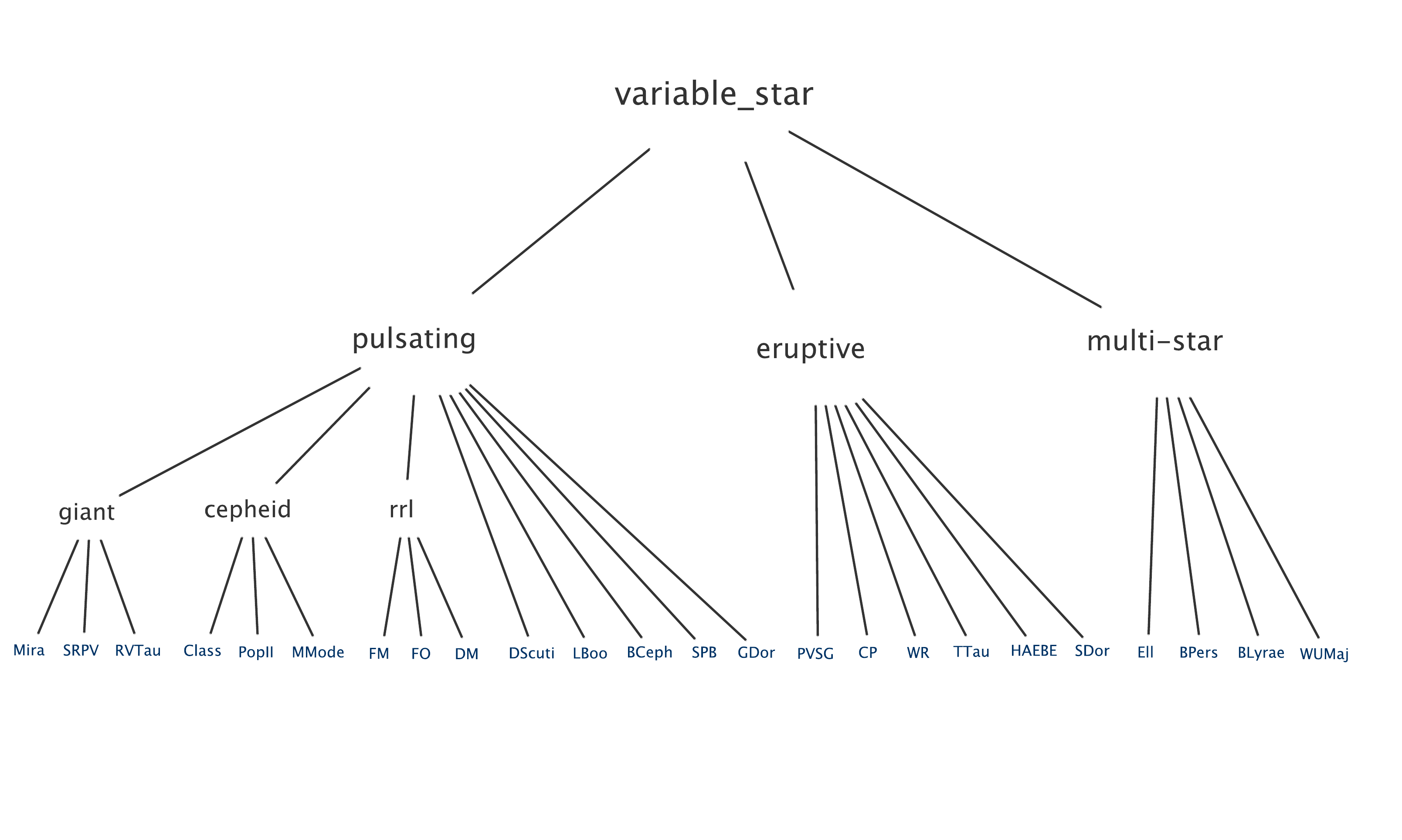}
\end{center}
\caption{Variable star classification hierarchy for the data used in \S\ref{sec:results}.  This structure is a crucial element of the two hierarchical classifiers used in this study, HSC and HMC.  The hierarchy is constructed based on knowledge of the physical processes that govern each type of object.  At the top level, the sources split into three major categories: pulsating, eruptive, and multi-star systems.  \label{fig:classhierarchy}}
\end{figure}

In variable star classification, we have at our disposal a well-established hierarchical taxonomy of classes based on the physics and phenomenology of these stars and stellar systems.  For instance, at the top level of our taxonomy, we can split the science classes into three main categories: pulsating, eruptive, and multi-star systems.  From there, we can continue to divide the sub-classes until we are left with exactly one of the original 25 science classes in each node (see figure \ref{fig:classhierarchy}).  For classification purposes, the meaning of the hierarchy is clear: mistakes at the highest levels of the hierarchy are more costly than mistakes made at deeper levels because the top levels of the hierarchy divide physical classes that are considerably different, whereas deeper levels divide subclasses that are quite similar.

Incorporating a known class hierarchy, such as that of figure \ref{fig:classhierarchy}, into a classification engine is a research field that has received much recent attention in the machine learning literature (see \citealt{2010sill} for a survey of these methods).  By considering the class hierarchy, these classifiers generally outperform their `flat classifier' counterparts because they impose higher penalties on the more egregious classification errors.  In this paper, we consider two types of hierarchical classification approaches: HSC (hierarchical single-label classification, \citealt{2006cesa}) and HMC (hierarchical multi-label classification, \citealt{2006bloc}).  We implement both HSC and HMC using random forests of decision trees.  Below, we provide a synopsis of HSC and HMC.   For more details about these methods, see \citet{2008vens}.  

In HSC, a separate classifier is trained at each non-terminal node in the class hierarchy, whereby the probabilities of each classifier are combined using conditional probability rules to obtain each of the class probabilities.  This has a similar flavor to the pairwise classifier approach of \S \ref{ss:pairwise}, but by adhering to the class hierarchy it need only build a small set of classifiers and can generate class probabilities in a straight-forward, coherent manner.  Moreover, different classifiers and/or sets of features can be used at each node in HSC, allowing for the use of more general classifiers at the top of the hierarchy and specialized domain-specific classifiers deeper in the hierarchy.  A recent paper of \citet{2010blom} applied a method similar to HSC, using Gaussian mixture classifiers, to classify variable stars observed by the {\it Kepler} satellite.  A second hierarchical classification approach is HMC, which builds a single classifier in which errors on the higher levels of the class hierarchy are penalized more heavily than errors deeper down the hierarchy.  In the version of HMC that we use, the weight given to a classification error at depth $d$ in the class hierarchy is $w_0^d$, where $w_0 \in (0,1)$ is a tuning parameter. This forces the algorithm to pay more attention to the top level, minimizing the instances of catastrophic error (defined in \S \ref{ss:comparison}). 

\subsection{Classifier assessment through cross-validation}
\label{ss:cv}

We have introduced a few methods that, given a sample of training data, estimate a classification model, $\widehat{p}$, to predict the science class of each new source.  In this section, we introduce statistically-rigorous methodology for assessing each classifier and choosing the optimal classifier amongst a set of alternatives.  Since our ultimate goal is to accurately classify newly-collected data, we will use the classifier that gives the best expected performance on new data.  We achieve this by defining a classifier's \emph{statistical risk} (i.e. prediction error) and computing an unbiased estimate of that risk via \emph{cross-validation}. We will ultimately use the model that obtains the smallest risk estimate.

Given a new source of class $Y$, having features $\mathbf{X}$, we define a loss function, $L(Y,\widehat{p}(\mathbf{X}))$, describing the penalty incurred by the application of our classifier, $\widehat{p}$, on that source.  The loss function encompasses our notion of how much the classifier $\widehat{p}$ has erred in predicting the source's class from its features $\mathbf{X}$.  The expected value of $L$, $E[L(Y,\widehat{p}(\mathbf{X}))]$, is the statistical risk, $R(\widehat{p})$, of the classifier.  The expected value, $E[\cdot]$, averages over all possible realizations of $(\mathbf{X},Y)$ to tell us how much loss we can expect to incur for the predicted classification of each new source (under the assumption that new data are drawn from the same distribution as the training data).  A key aspect to this approach is that it guards against over-fitting to the training data: if a model is overly complex, it will only add extra variability in classification without decreasing the bias of the classifier, leading to an increase in the risk (this is the bias-variance tradeoff, see \citealt{2006wass}).

Conveniently, within this framework each scientist is free to tailor the loss function to meet their own scientific goals.  For instance, an astronomer interested in finding Mira variables could define a loss function that incurs higher values for mis-classified Miras.  In this work, we use the vanilla 0-1 loss that is defined to be 0 if $Y = \widehat{p}(\mathbf{X})$ and 1 if misclassified (here, $\widehat{p}(\mathbf{X})$ is the estimated class of the source; alternatively, we could define a loss function over the set of estimated class probabilities, $\{\widehat{p}_1(\mathbf{X}),...,\widehat{p}_C(\mathbf{X})\}$).  Under 0-1 loss, the statistical risk of a classifier is its expected overall misclassification rate, which we aim to minimize.

There remains the problem of how to estimate the statistical risk, $R(\widehat{p})$ of a classifier $\widehat{p}$.  If labeled data were plentiful, we could randomly split the sources into training and validation sets, estimating $\widehat{p}$ with the training set and computing a risk estimate $\widehat{R}(\widehat{p})$ with the validation set.  Since our data are relatively small in number, we use $k$-fold cross-validation to estimate $R$.  In this procedure, we first split the data into $K$ (relatively) equal-sized parts.  For each subset $k=1,...,K$ the classifier is fitted on all of the data not in $k$ and a risk estimate, $\widehat{R}_{(-k)}(\widehat{p})$, is computed on the data in set $k$.  The cross-validation risk estimate is  defined as $\widehat{R}_{\rm CV}(\widehat{p})= \frac{1}{K} \sum_{k=1}^K\widehat{R}_{(-k)}(\widehat{p})$.  As shown in \citet{1989burm}, for $K \gtrsim 4$, $\widehat{R}_{\rm CV}(\widehat{p})$ is an approximately-unbiased estimate of $R(\widehat{p})$.  In this paper, we use $K=10$ fold cross-validation.

In addition to selecting between different classification models, cross-validation risk estimates can be used to choose the appropriate tuning parameter(s) for each classifier.  For instance, we use cross-validation to choose the optimal pruning depth for classification trees, to pick the optimal number and depth of trees to build, and number of candidate splitting features to use at each split for random forests, and to select the optimal size of the base learner, number of trees, and learning rate for boosted decision trees.  The optimal set of tuning parameters for each method is found via a grid search.  For each of the methods considered, we find that $\widehat{R}_{\rm CV}$ is stable in the neighborhood of the optimal tuning parameters, signifying that the classifiers are relatively robust to the specific choice of tuning parameters and that a grid search over those parameters is sufficient to obtain near-optimal results.\\

\section{Classifier performance on OGLE+Hipparcos data set}
\label{sec:results}

\subsection{Description of data}
\label{sec:data}

In this paper, we test our feature extraction and classification methods using a mixture of variable star photometric data from the OGLE and Hipparcos surveys.  OGLE (Optical Gravitational Lensing Experiment, \citealt{1999udal}) is a ground-based survey from Las Campanas Observatory covering fields in the Magellanic Clouds and Galactic bulge.  Hipparcos  Space Astrometry Mission (\citealt{1997perr}) was an ESA project designed to precisely measure the positions of more than one hundred thousand stars.  The data selected for this paper are the OGLE and Hipparcos sources analyzed by \citet{2007debo}, totaling 90\% of the variable stars studied in that paper.  A summary of the properties, by survey, of the data used in our study, is in table \ref{tab:surveycharacs}. The light curve data and classifications used for each source can be obtained through our {\tt dotastro.org} light curve repository.\footnote{\url{http://dotastro.org/lightcurves/project.php?Project\_ID=123}}

This sample was designed by \citet{2007debo} to provide a sizable set of stars within each science class, for a broad range of classes.  Our sample contains stars from the 25 science classes analyzed in their paper.  Via an extensive literature search, they obtained a set of confirmed stars of each variability class.  In table \ref{tab:sciclasscharacs} we list, by science class, the proportion of stars in that data set that we have available for this paper.  Since the idea of their study was to capture and quantify the typical variability of each science class, the light curves were pre-selected to be of good quality and to have an adequate temporal sampling for accurate characterization of each science class.  For example, the multi-mode Cepheid and double-mode RR Lyrae stars, which have more complicated periodic variability, were sampled from OGLE because of its higher sampling rate.

In our sample, there are 25 objects that are labeled as two different science classes.  Based on a literature search of these stars, we determine that 14 of them reasonably belong to just a single class (5 S Doradus, 2 Herbig AE/BE, 3 Wolf Rayet, 2 Delta Scuti, 1 Mira, and 1 Lambda Bootis).  The other 11 doubly-labeled stars, which are listed in table \ref{tab:throwouts},  were of an ambiguous class or truly belonged to two different classes, and were removed from the sample.  See Appendix \ref{app:double} for a detailed analysis and references for the doubly-labeled objects. Because the sample was originally constructed by \citet{2007debo} to consist only of well-understood stars with confident class labeling, we are justified in excluding these sources.\\

\begin{deluxetable}{lccccc} 
\tablecolumns{6} 
\tablewidth{0pc} 
\tablecaption{Dataset characteristics, by survey.} 
\tablehead{ 
\colhead{Survey} & \colhead{NLC\tablenotemark{a}} & \colhead{$\% {\rm NLC}_{deb}$\tablenotemark{b}} & \colhead{NLC used\tablenotemark{c}} & \colhead{$\langle T_{\rm tot} \rangle$ (days)\tablenotemark{d}} & \colhead{$\langle N_{\rm epochs}\rangle$} }
\startdata 
HIPPARCOS	 &1044	 &100.0 & 1019 &1097	 &103\\ 
OGLE	 &523	 &99.2 & 523 &1067 &329
\enddata
\tablenotetext{a}{Total number of light curves available to us.}
\tablenotetext{b}{Percentage of \citet{2007debo} light curves available to us.}
\tablenotetext{c}{Number of light curves after removal of sources with ambiguous class and exclusion of small classes.}
\tablenotetext{d}{Average time baseline.}
\label{tab:surveycharacs}
\end{deluxetable}%

\begin{deluxetable}{llccccccc} 
\tablecolumns{9} 
\tablewidth{0pc} 
\tablecaption{Dataset characteristics, by science class.} 
\tablehead{ 
\colhead{Variable Star Class}	 & \colhead{${\rm Name}_{\rm deb}$\tablenotemark{a}}	 & \colhead{NLC\tablenotemark{b}} & \colhead{$\% {\rm NLC}_{deb}$} & \colhead{Instrument} & \colhead{$\langle N_{\rm epochs} \rangle$} & \colhead{$min(f_1)$\tablenotemark{c}} & \colhead{$\langle f_1 \rangle$\tablenotemark{c}} & \colhead{$max(f_1)$\tablenotemark{c}}}
\startdata 
a. Mira & 				MIRA	 &144	 &100.0		&HIPPARCOS&98	 &0.0020	&0.09	 &11.2508\\
 b. Semireg PV &		SR	 &42	 &100.0		&HIPPARCOS&99	 &0.0010	&0.15	 &1.0462\\
c. RV Tauri &			RVTAU	 &6	 &46.2		&HIPPARCOS&104	 &0.0012	&0.05	 &0.1711\\
d. Classical Cepheid&	CLCEP	 &191	 &97.9		&HIPPARCOS&108	 &0.0223	&0.15	 &0.4954\\
e. Pop. II Cepheid&		PTCEP	 &23	 &95.8		&HIPPARCOS&107	 &0.0037	&0.21	 &0.7648\\
f. Multi. Mode Cepheid &	DMCEP	 &94	 &98.9		&OGLE&181	 &0.5836	&1.21	 &1.7756\\
 g. RR Lyrae, FM &		RRAB	 &124	 &96.1		&HIPPARCOS&91	 &1.2149	&1.95	 &9.6197\\
h. RR Lyrae, FO  &		RRC	 &25	 &86.2		&HIPPARCOS&92	 &2.2289	&3.15	 &4.3328\\
i. RR Lyrae, DM  &		RRD	 &57	 &100.0		&OGLE&304	 &2.0397	&2.61	 &2.8177\\
 j. Delta Scuti &			DSCUT	 &114	 &82.0		&HIPPARCOS&129	 &0.0044	&7.90	 &19.7417\\
k. Lambda Bootis &		LBOO	 &13	 &100.0		&HIPPARCOS&84	 &7.0864	&12.36	 &19.8979\\
 l. Beta Cephei &		BCEP	 &39	 &67.2		&HIPPARCOS&96	 &0.0014	&4.94	 &10.8319\\
m. Slowly Puls. B &		SPB	 &29	 &61.7		&HIPPARCOS&101	 &0.1392	&1.09	 &11.8302\\
n. Gamma Doradus &	GDOR	 &28	 &80.0		&HIPPARCOS&95	 &0.2239	&2.24	 &9.7463\\
o. Pulsating Be & 		BE	 &45	 &78.9		&HIPPARCOS&106	 &0.0011	&2.12	 &14.0196\\
p. Per. Var. SG &		PVSG	 &55	 &72.4		&HIPPARCOS&102	 &0.0015	&3.41	 &15.7919\\
q. Chem. Peculiar&		CP	 &51	 &81.0		&HIPPARCOS&105	 &0.0076	&2.57	 &13.4831\\
r. Wolf-Rayet &			WR	 &41	 &65.1		&HIPPARCOS&99	 &0.0011	&6.56	 &19.2920\\
 s. T Tauri &			TTAU	 &14	 &82.4		&HIPPARCOS&67	 &0.0013	&1.85	 &11.2948\\
t. Herbig AE/BE &		HAEBE	 &15	 &71.4		&HIPPARCOS&83	 &0.0009	&1.41	 &10.0520\\
u. S Doradus &			LBV	 &7	 &33.3		&HIPPARCOS&95	 &0.0008	&0.20	 &0.5327\\
v. Ellipsoidal &			ELL	 &13	 &81.2		&HIPPARCOS&105	 &0.1070	&1.37	 &3.5003\\
w. Beta Persei  &		EA	 &169	 &100.0		&OGLE&375	 &0.0127	&0.93	 &3.1006\\
  x. Beta Lyrae  &		EB	 &145	 &98.6		&OGLE&365	 &0.0175	&0.71	 &4.5895\\
y. W Ursae Maj. &		EW	 &58	 &98.3		&OGLE&369	 &0.2232	&2.44	 &8.3018
\enddata
\tablenotetext{a}{Class name in \citet{2007debo}.}
\tablenotetext{b}{Total number of light curves used, after removal of ambiguous sources.}
\tablenotetext{c}{$f_1$ is the frequency of the first harmonic in ${\rm day}^{\rm -1}$.}
\label{tab:sciclasscharacs}
\end{deluxetable}

\subsection{Class-wise distribution of light-curve features}

Using the methodology in \S \ref{sec:features}, we estimate features for each variable star in the data set using their light curve.  The feature-extraction routines take 0.8 seconds per light curve, giving us a 53-dimensional representation of each variable star.  The computations are performed in {\tt Python} and {\tt C} using a non-parallelized, single thread on a 2.67 GHz Intel Xeon X5550 CPU running on a v2.6.18 linux kernel machine.  We estimate that the periodic-feature routines account for 75\% of the computing time and scale linearly with the number of epochs in the light curve.  Note that these metrics do not take into account the CPU time needed to read the XML data files from disk and load the data into memory.

\begin{figure}
\begin{center}
$\begin{array}{c@{\hspace{.1in}}c}
\multicolumn{1}{l}{\mbox{\bf (a)}} &	\multicolumn{1}{l}{\mbox{\bf (b)}} \\ [-.35in]
\includegraphics[angle=0,width=3.2in]{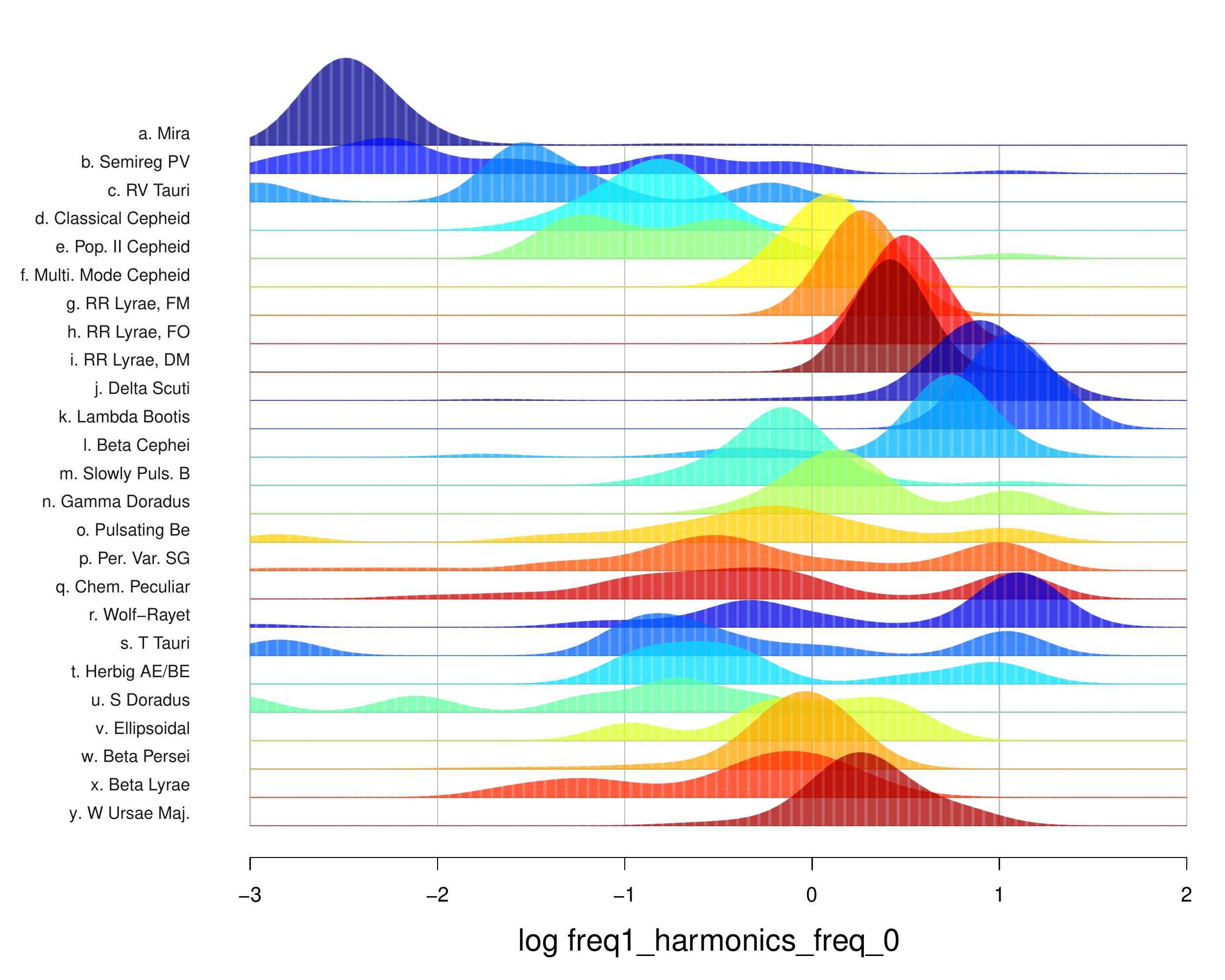}&
\includegraphics[angle=0,width=3.2in]{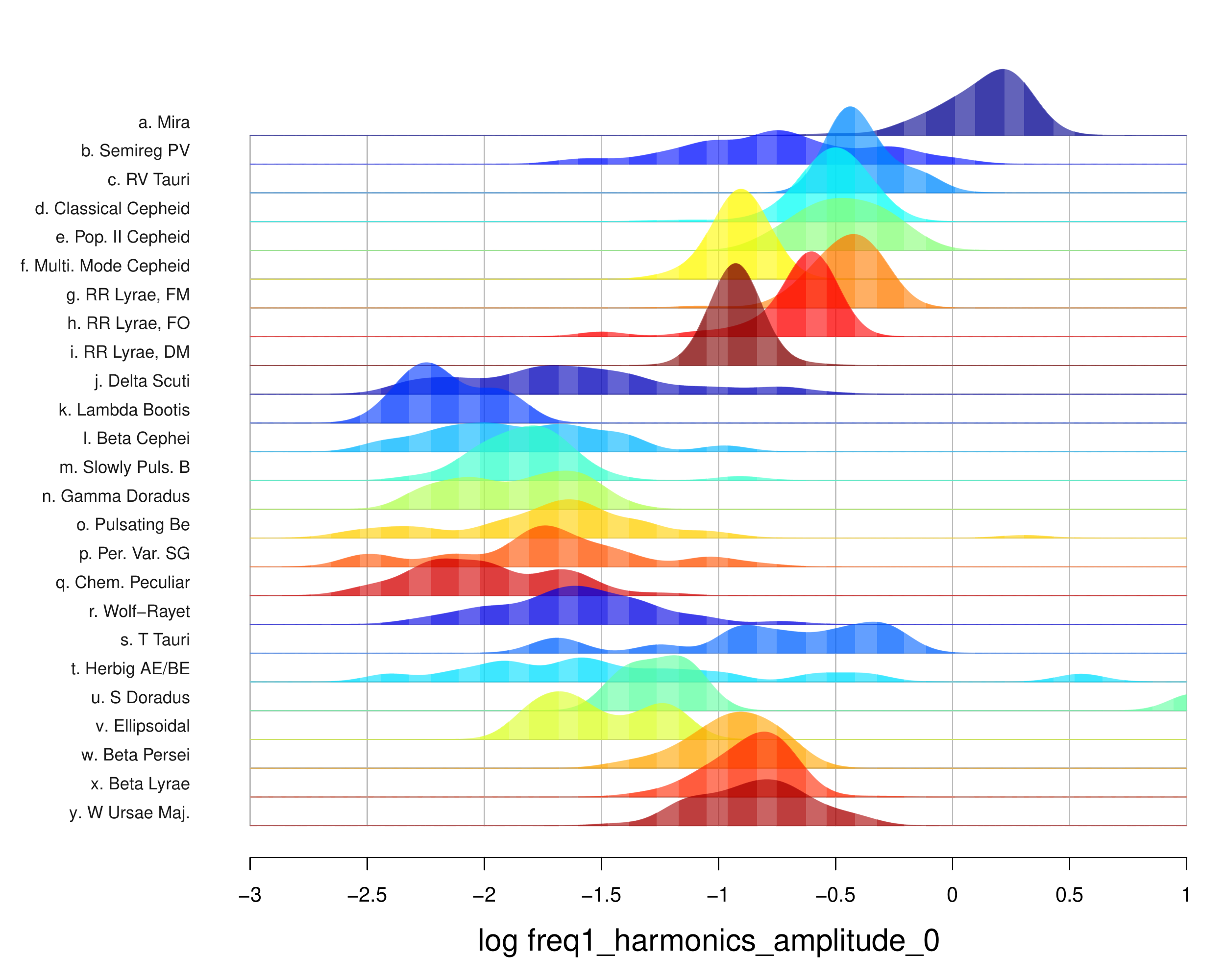}\\ 
\multicolumn{1}{l}{\mbox{\bf (c)}} &	\multicolumn{1}{l}{\mbox{\bf (d)}} \\ [-.35in]
\includegraphics[angle=0,width=3.2in]{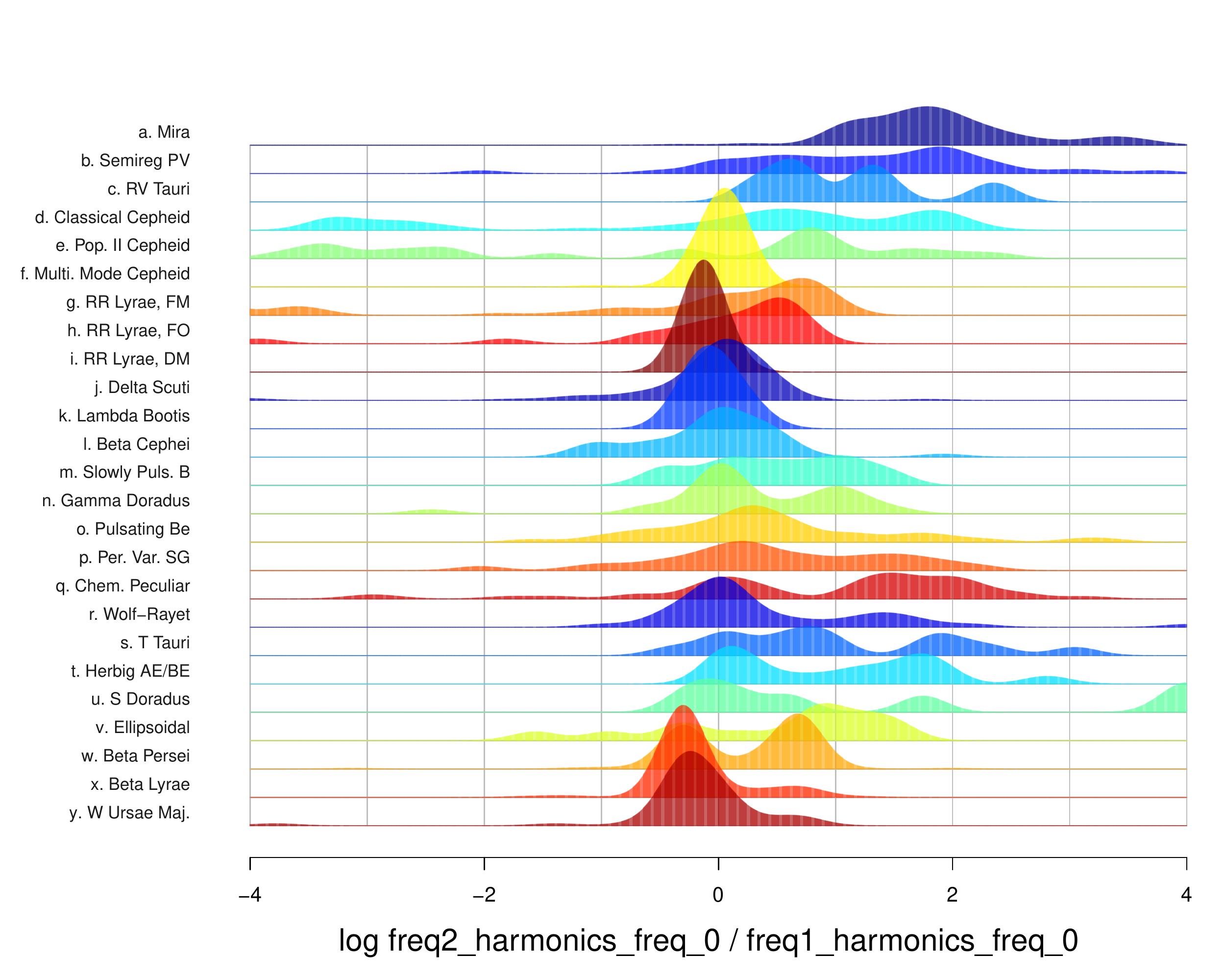}&
\includegraphics[angle=0,width=3.2in]{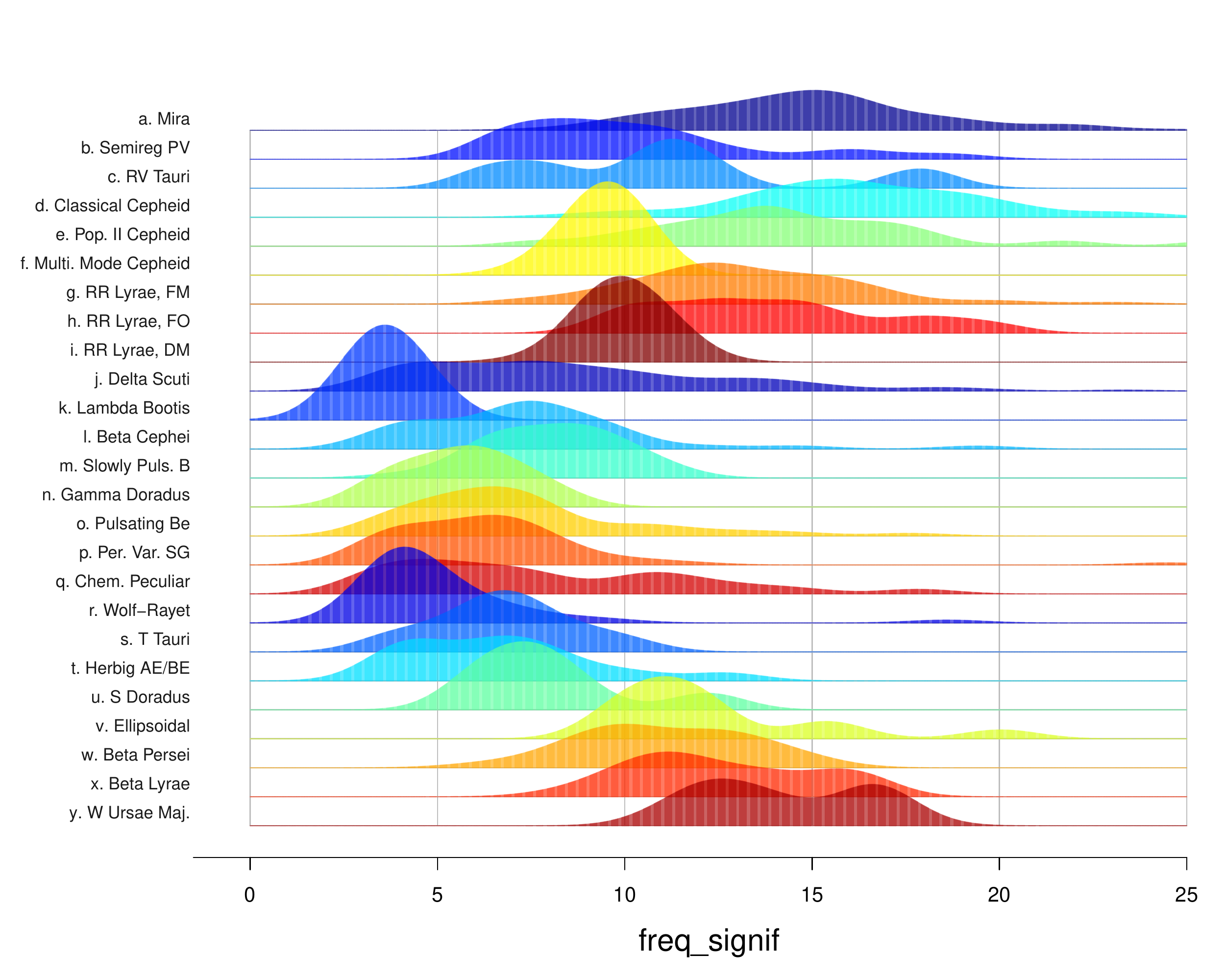}\\
\multicolumn{1}{l}{\mbox{\bf (e)}} &	\multicolumn{1}{l}{\mbox{\bf (f)}} \\ [-.35in]
\includegraphics[angle=0,width=3.2in]{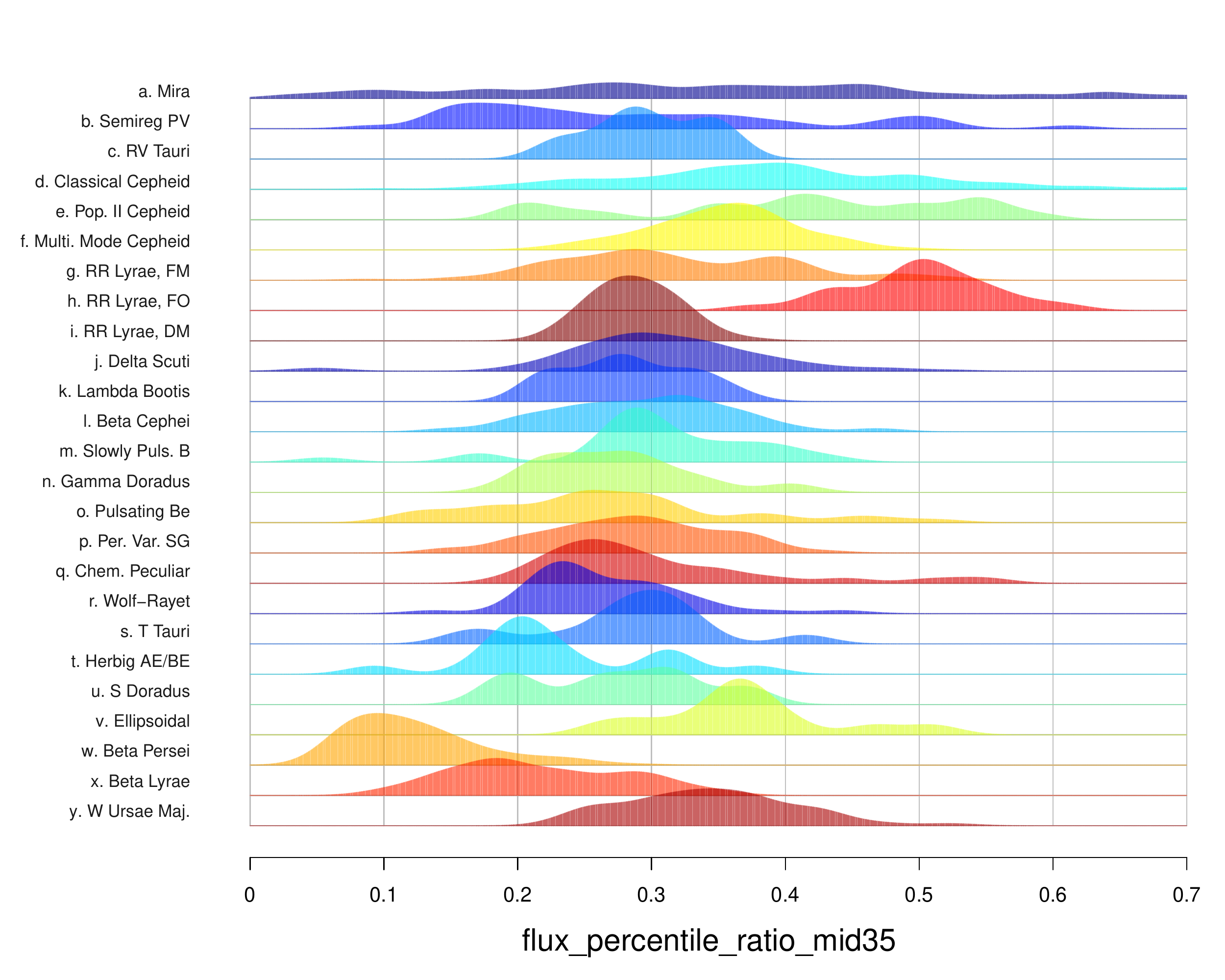}&
\includegraphics[angle=0,width=3.2in]{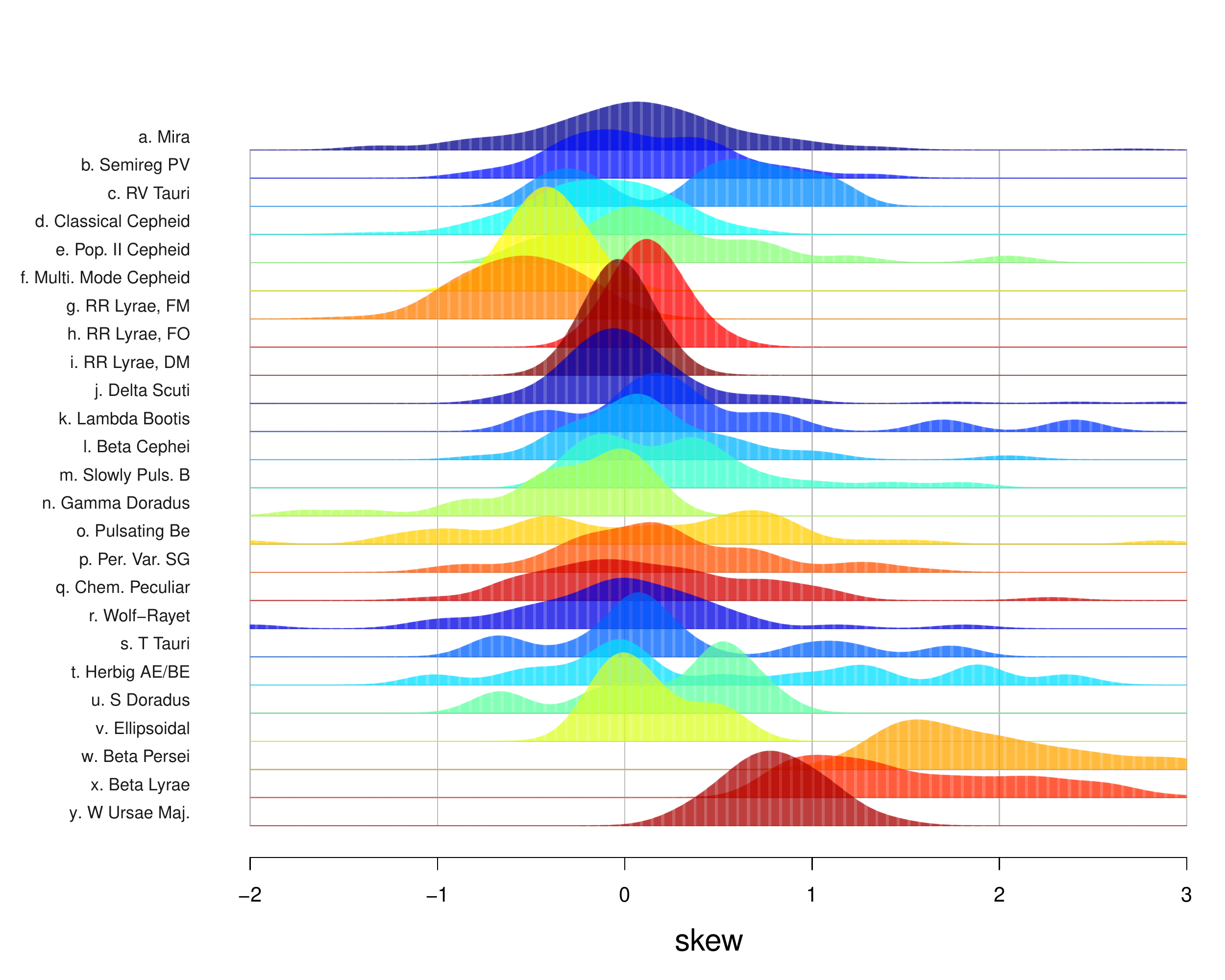}\\
\multicolumn{1}{l}{\mbox{\bf (g)}} &	\multicolumn{1}{l}{\mbox{\bf (h)}} \\ [-.35in]
\includegraphics[angle=0,width=3.2in]{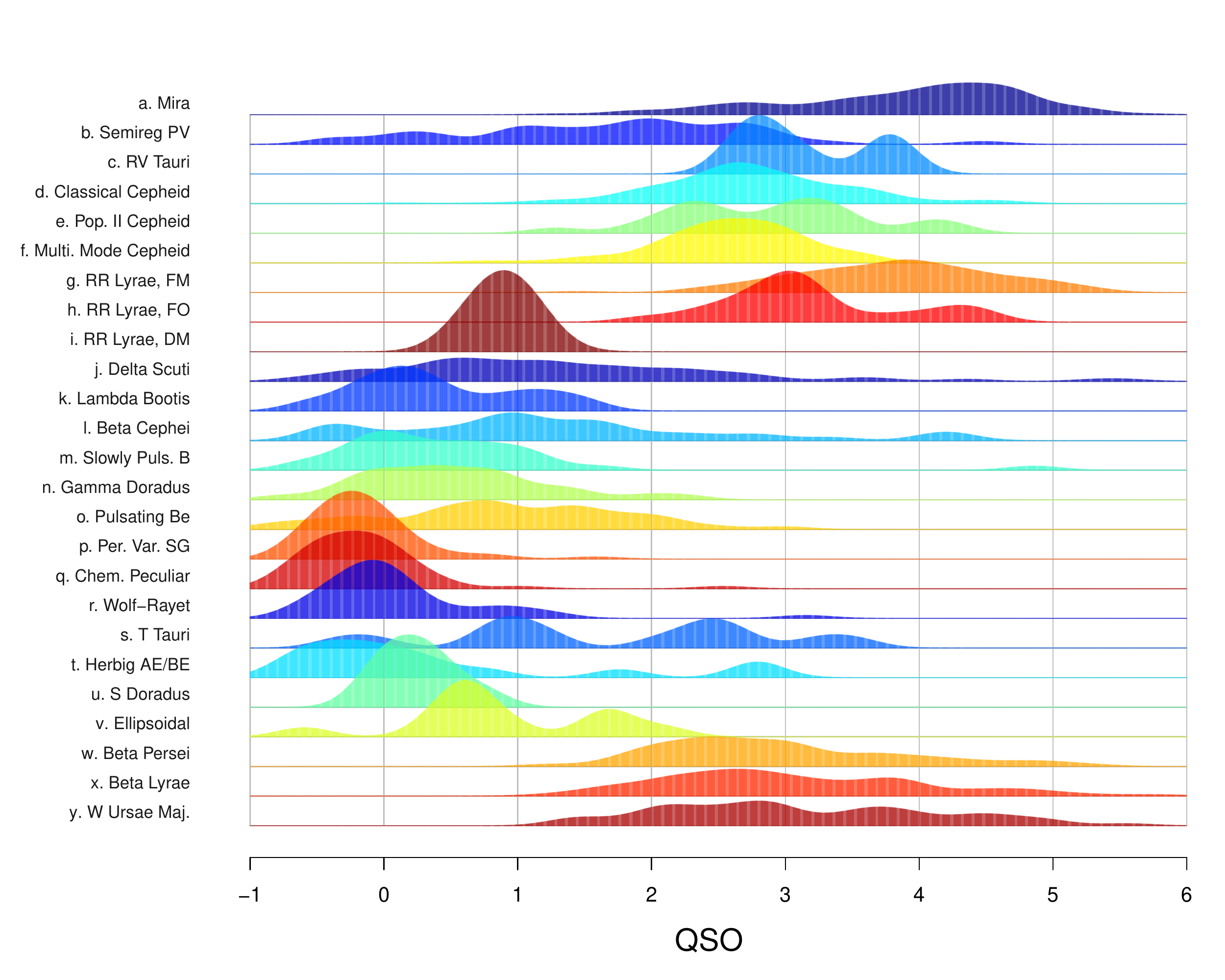}&
\includegraphics[angle=0,width=3.2in]{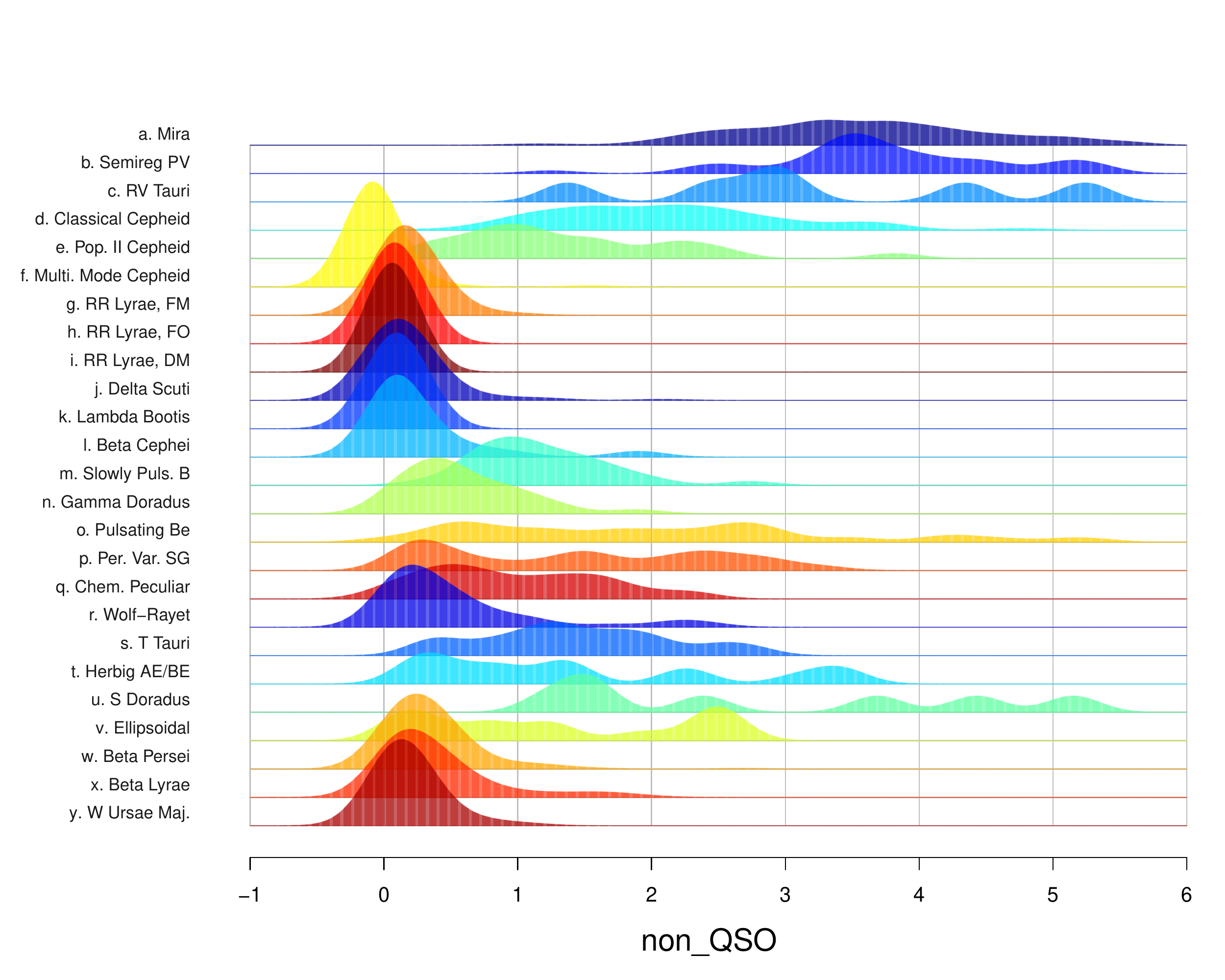}\\[-.75cm]
\end{array}$
\end{center}
\caption{Histograms of several features, by class.  The features plotted are (a) the first frequency in cycles/day, (b) amplitude of the first frequency in mag, (c) the ratio of the second to the first frequencies, (d) statistical significance of the periodic model, (e) flux ratio middle 35th to middle 95th quantiles, (f) flux skew, (g) \citealt{2010butl} QSO variability, and (h) \citealt{2010butl} non-QSO variability.  \label{fig:LSfeat}}
\end{figure}

Plots of 1-dimensional density estimates, by science class, of a selected set of features are in figure \ref{fig:LSfeat}.  These class-wise feature distributions allow us to quickly and easily identify the differences, in feature space, between individual variable star science classes.  Density plots are very useful for this visualization because they provide a complete feature-by-feature characterization of each class, showing any multi-modality, outliers, and skewness in the feature distributions.  For instance, it is immediately obvious that several of the eruptive-type variable star classes have an apparent bi-modal or relatively flat frequency distributions, likely attributed to their episodic nature.  Conversely, the RR Lyrae frequency distributions are all narrow and peaked, showing that indeed these stars are well-characterized by the frequency of their flux oscillations.  The feature density plots also inform us of which feature(s) are important in separating different sets of classes.  For example, the RR Lyrae, FO and RR Lyrae, DM stars have overlapping distributions for each of the features in figure \ref{fig:LSfeat} except the feature QSO, where their distributions are far apart, meaning that QSO will be a useful classification feature in separating stars of those two classes.

\subsection{Classifier comparison}
\label{ss:comparison}

In this section, we compare the different classification methods introduced in \S\ref{sec:methods}.  To fit each classifier, except HMC-RF, we use the statistical environment {\tt R}\footnote{{\tt R} is a freely-available language and environment for statistical computing and graphics available at {\tt http://cran.r-project.org/}}.  To fit HMC-RF we use the open-source decision tree and rule learning system {\tt Clus}.\footnote{{\tt http://dtai.cs.kuleuven.be/clus/index.html}}  Each of the classifiers was tuned via a grid search over the relevant tuning parameters to minimize the cross-validation misclassification rates.  To evaluate each classifier, we consider two separate metrics: the overall misclassification error rate and the catastrophic error rate.  We define catastrophic errors to be any classification mistake in the top-level of the variable-star hierarchy in figure \ref{fig:classhierarchy} (i.e., pulsating, eruptive, multi-star).  The performance measures for each classifier, averaged over 10 cross-validation trials, are listed in table \ref{tab:misclass}.  In terms of overall misclassification rate, the best classifier is a random forest with $B=1000$ trees, achieving a 22.8\% average misclassification rate.  In terms of catastrophic error rate, the HSC-RF classifier with $B=1000$ trees achieves the lowest value, 7.8\%.  The processing time required to fit the ensemble methods is greater than the time needed to fit single tree models.  However, this should not be viewed as a limiting factor: once any of these models is fit, predictions for new data can be produced very rapidly.  For example, for a random forest classifier of 1000 trees, class probability estimates can be generated for new data at the rate of 3700 instances per second.

\begin{deluxetable}{lccc} 
\tablecolumns{3} 
\tablewidth{0pc} 
\tablecaption{Performance of Classifiers on OGLE+Hipparcos data set,  averaged over 10 repetitions.} 
\tablehead{ 
\colhead{Method} & \colhead{Misclassification \%\tablenotemark{a}} & \colhead{Catastrophic error \%\tablenotemark{a}}   &  \colhead{CPU\tablenotemark{b}}   }
\startdata 
 CART&     32.2 &13.7&  10.6\\
 C4.5    & 29.8 &12.7  &14.4\\
 RF      & {\bf 22.8}  &8.0 &117.6\\
 Boost   & 25.0 & 9.9 &466.5\\
 CART.pw&  25.8 & 8.7& 323.2\\
 RF.pw   & 23.4  &8.0 &290.3\\
 Boost.pw &24.1 & 8.2 &301.5\\
 SVM.pw   &25.3 & 8.4 &273.0\\
 HSC-RF   &23.5 & {\bf 7.8} &230.9\\
 HMC-RF   &23.4&  8.2 &946.0
\enddata
\tablenotetext{a}{Estimated using 10-fold cross-validation.}
\tablenotetext{b}{Average processing time in seconds for 10-fold cross-validation on a 2.67 GHz Macintosh with 4 GB of RAM.}
\label{tab:misclass}
\end{deluxetable}%

\subsubsection{Tree-based classifiers}

On average, the single-tree classifiers--CART and C4.5--are outperformed by the tree ensemble methods--random forest and tree boosting--by 21\% in terms of misclassification error rate.  The classification performances of the single-tree classifiers are near that of the best classifier in \citet{2007debo}, which achieves a 30\% error rate. Tree-based classifiers seem particularly adept at variable-star classification, and ensembles of trees achieve impressive results.  Single trees take on the order of 10 seconds to fit a model, prune based on cross-validation complexity, and predict classification labels for test cases. Tree ensemble methods take 10-40 times longer to fit 1000 trees. Overall, the random forest classifier is the best classification method for these data: it achieves the lowest overall misclassification rate, low catastrophic error rate, and is the third fastest algorithm.

\subsubsection{Pairwise classifiers}

We implement four pairwise classifiers: CART, random forest, boosted trees, and SVM.  Of these, random forest achieves the best results in terms of both misclassification rate and catastrophic error rate, at 23.4\% and 8.0\%, respectively.  The pairwise classifiers all perform better than single-tree classifiers but tend to fare worse than the single random forest method.  It is interesting to note that our implementation of SVM achieves a 25.3\% misclassification rate, a vast improvement over the 50\% SVM misclassification rate found by \citet{2007debo}.  This is likely due to both our use of better features and our pre-selection of the 25 features (chosen via cross-validation) with highest random-forest variable importance for use in the SVM.  Unlike tree models, SVM is not immune to the inclusion of many useless features; when we include all 53 features into the SVM, our error rate skyrockets to 54\%.

\subsubsection{Hierarchical classifiers}

Two of our classifiers, HSC-RF and HMC-RF, incorporate the hierarchical class taxonomy when building a classifier.  Both of these methods achieve overall classification error rates slightly worse (sub-1\% level) than that of the random forest classifier, while HSC-RF reaches the best overall catastrophic error rate (7.8\%).  HMC-RF slightly outperforms HSC-RF with respect to misclassification rate, but its current implementation takes 4 times as much CPU time.  HSC-RF, HMC-RF, pairwise random forest and the original random forest are the best methods in terms of error rates, but random forest is at least twice as fast as any of the other methods.

\subsection{Direct comparison to \citet{2007debo}}

Random forest achieves a 22.8\% average misclassification rate, a  24\% improvement over the 30\% misclassification rate achieved by the best method of \citet{2007debo} (Bayesian model averaging of artificial neural networks).  Furthermore, each of the classifiers proposed in this paper, except the single-tree models CART and C4.5, achieves an average misclassification rate smaller than 25.8\% (see table \ref{tab:misclass}).  There is no large discrepancy between the different ensemble methods---the difference between the best (random forest) and worst (boosting) ensemble classifier is $2.2\%$, or an average of 34 more correct classifications---but in terms of both accuracy and speed, random forest is the clear winner.

\begin{figure}
\epsscale{.8}
\begin{center}
\includegraphics[angle=0,scale=.5]{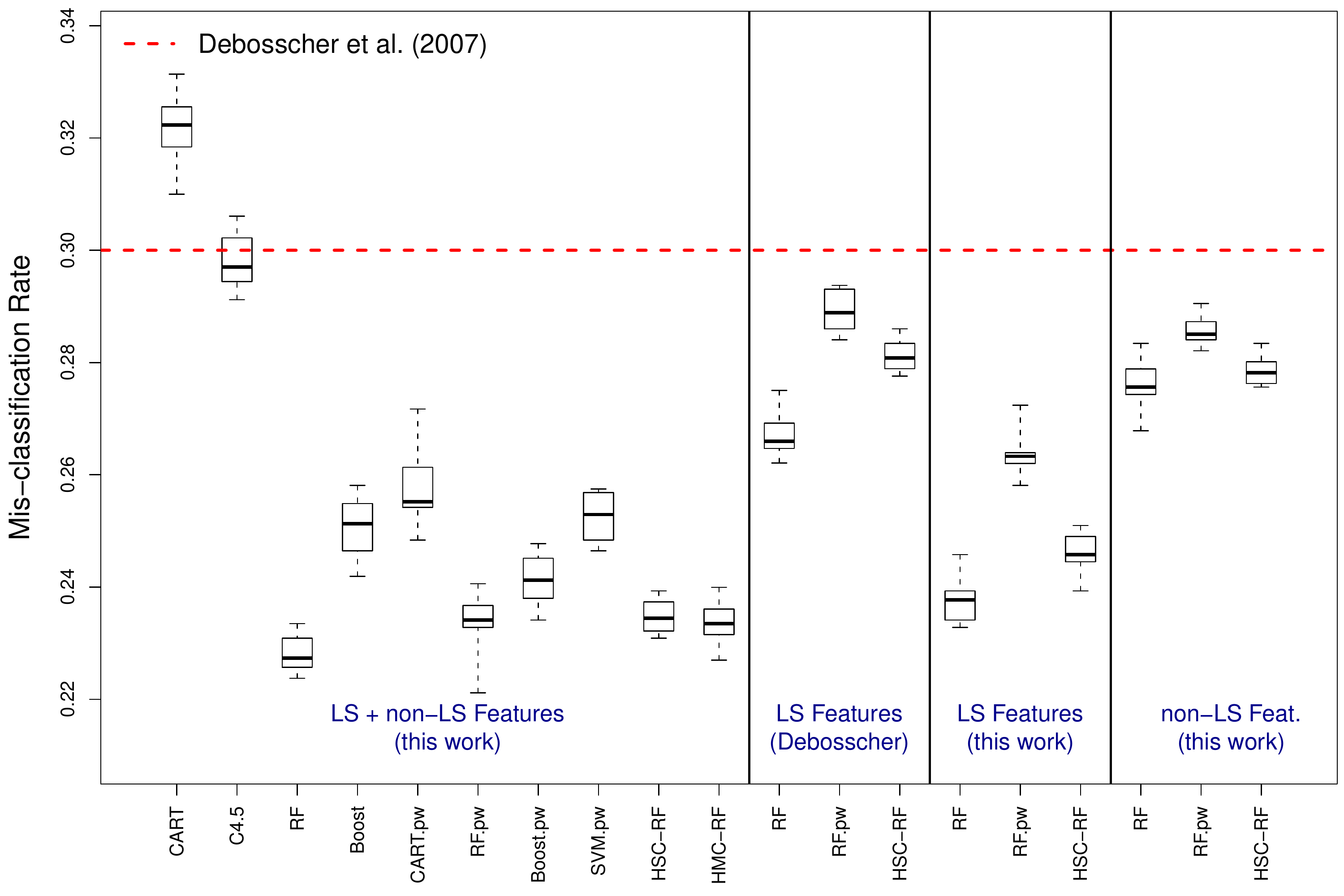}
\end{center}
\caption{Distribution of cross-validation error rates for several classifiers on the OGLE+Hipparcos data, obtained from 10 repetitions.  The classifiers are divided based on the features on which they were trained; from left to right: (1) all of the periodic and non-periodic features that we estimate, (2) the Lomb-Scargle features estimated by \citet{2007debo}, (3) the Lomb-Scargle features estimated by us, and (4) the non-periodic features.  In terms of mis-classification rate, the classifiers trained on all of our features perform the best, followed by those trained only on our periodic features, those trained on Debosscher et al.'s periodic features, and those trained on only the non-periodic features.  All of the classifiers used, except single trees, achieve better error rates than the best classifier from Debosscher et al. (dashed line). \label{fig:misclass}}
\end{figure}

In comparing our results to the \citet{2007debo} classification results, it is useful to know whether the gains in accuracy are due to the use of better classifiers, more accurate periodic feature extraction, informative non-periodic features, or some combination of these.  To this end, we classify these data using the following sets of features:
\begin{enumerate}
\item The periodic features estimated by \citet{2007debo}.
\item Our estimates of the periodic features following \S \ref{sec:features}.
\item The non-periodic features proposed in \S \ref{sec:features} 
\item The non-periodic features in addition to our periodic features.
\end{enumerate}
Misclassification rates from the application of our classifiers to the above sets of features are plotted in figure \ref{fig:misclass}.  As a general trend, using both our periodic and non-periodic features is better than using only our periodic features, which is in turn better than using Debosscher et al.'s periodic features, which achieves similar rates to using only the non-periodic features.  Using a random forest classifier, we find that the average cross-validated misclassification rates are 22.8\% using all features, 23.8\% using our periodic features, 26.7\% using \citet{2007debo} features, and 27.6\% using only our non-periodic features.  This is evidence that we obtain better classification results both because our classification model is better and because the extracted features we use are more informative.


\subsection{Analysis of classification errors}

\begin{figure}
\epsscale{.8}
\includegraphics[angle=0,scale=.8]{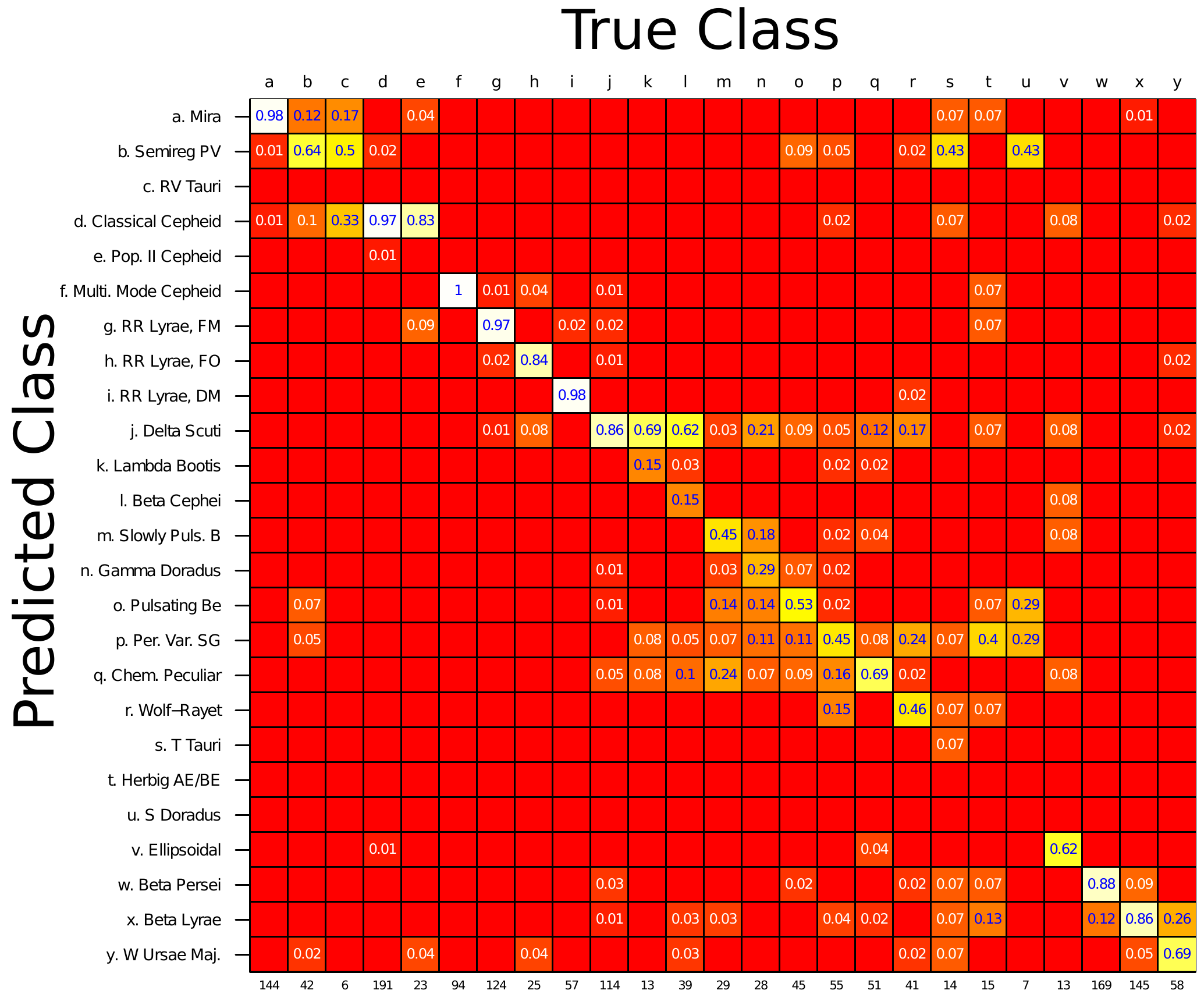}
\caption{Cross-validated confusion matrix obtained by the random forest classifier for the OGLE+Hipparcos data.  A perfect classifier would place all mass on the diagonal.  We have sorted the science classes by physical similarity, so large numbers near the diagonal signify that the classifier  is performing well.  On average, the classifier performs worse on the eruptive events, as exemplified by a large spread of mass for classes p through u.  The overall error rate for this classifier was 21.1\%.\label{fig:confmat}}
\end{figure}

Understanding the source and nature of the mistakes that our classifier makes can alert to possible limitations in the classification methods and feature extraction software and aid in the construction of a second-generation light-curve classifier.  Let us focus on the classifier with best overall performance: the random forest, whose cross-validated misclassification rate was 22.8\%.  In figure \ref{fig:confmat} we plot the confusion matrix of this classifier, which is a tabular representation of the classifier's predicted class versus the true class.  A perfect classifier would place all of the data on the diagonal of the confusion matrix; any deviations from the diagonal inform us of the types of errors that the classifier makes. 

A few common errors are apparent in figure \ref{fig:confmat}.  For instance, Lambda Bootis and Beta Cephei are frequently classified as Delta Scuti.  Wolf-Rayet, Herbig AE/BE and S Doradus stars are usually classified as Periodic Variable Super Giants, and T Tauri are often misclassified as Semiregular Pulsating Variables.  None of these mistakes are particularly egregious: often the confused science classes are physically-similar.  Also, the mis-classified examples often come from classes with few training examples (see below) or characteristically low-amplitude objects (see \S \ref{sec:highamp}).

\begin{figure}
\begin{center}
\epsscale{.8}
\includegraphics[angle=0,scale=.6]{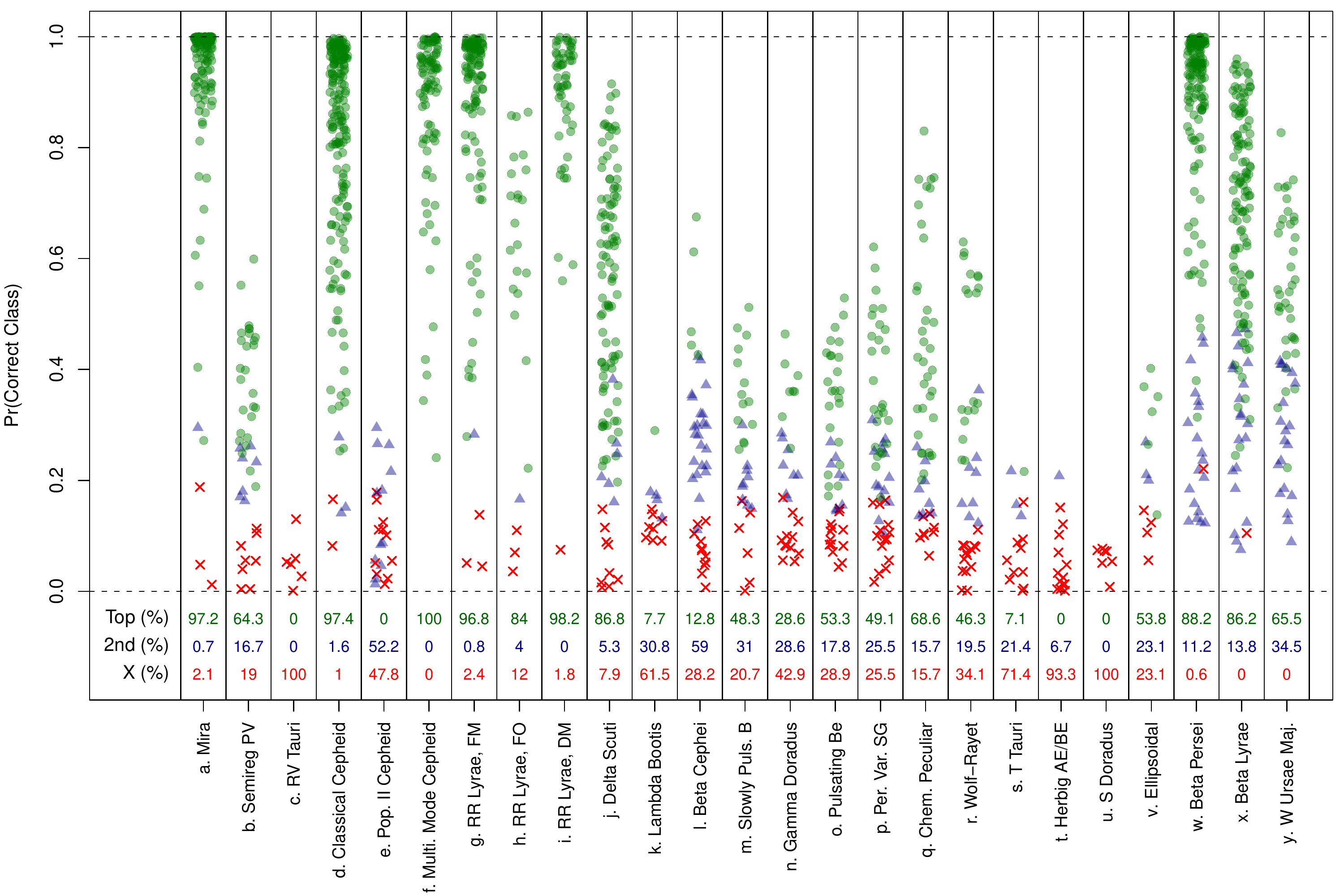}
\end{center}
\caption{Cross-validated random forest probability estimate of the correct class for each variable star, plotted by class.  Green circles indicate that the classifier's top choice for that object was correct, blue triangles indicate that the second choice was correct, and red x's indicated that neither of the top two were correct.  As a general trend, we find that the classifier performs better for large classes. \label{fig:clProbs}}
\end{figure}

As the random forest is a probabilistic classifier, for each source it supplies us with a probability estimate that the source belongs to each of the science classes.  Until now, we have collapsed each vector of probabilities into an indicator of most probable class, but there is much information available to extract from the individual probabilities.  For instance, in figure \ref{fig:clProbs} we plot, by class, the random-forest estimated probability that each source is of its true class.  We immediately see a large disparity in performance between the classes: for a few classes, we estimate high probabilities of true class, while for others we generally estimate low probabilities.  This discrepancy is related to the size of each class: within the science classes that are data-rich, we tend to get the correct class, while in classes with scarce data, we usually estimate the wrong class.  This same effect is seen in \citet{2007debo}, (their table 5).  This is a common problem in statistical classification for unbalanced class sizes: classifiers such as random forest try to minimize the overall classification rate, thus focusing most of their efforts on the larger classes.  One can remedy this by weighting the training data inversely proportional to their class size.  However, the price to pay for the increase in balanced error among the classes is a higher overall misclassification rate (see \citealt{1984brei}).  The better solution is obviously to obtain more training examples for the under-sampled classes to achieve a better characterization of these classes.

We have experimented with a random forest classifier using inverse class-size weighting.  The results of this experiment were as expected: our overall misclassification rate climbs to 28.0\%, a 23\% increase in error over the standard random forest, but we perform better within the smaller classes.  Notably, the number of correctly-classified Lambda Bootis increases from 1 to 7, while the number of correctly-classified Ellipsoidal variables jumps from 6 to 11, Beta Cephei from 5 to 23, and Gamma Doradus from 8 to 15.  All four classes in which the original random forest found no correct classifications each had at least two correct classifications with the weighted random forest.

\subsection{Performance for specific science classes}

Although our classifier was constructed to minimize the number of overall misclassifications in the 25-class problem, we can also use it to obtain samples of objects from science classes of interest via probability thresholding.  The random forest classifier produces a set of class-wise posterior probability estimates for each object.  To construct samples of a particular science class, we define a threshold on the posterior probabilities, whereby any object with class probability estimate larger than the threshold is included in the sample.  By decreasing the threshold, we trade off purity of the sample for completeness, thereby mapping out the receiver operating characteristic (ROC) curve of the classifier.

\begin{figure}
\begin{center}
$\begin{array}{c@{\hspace{.1in}}c}
\multicolumn{1}{l}{\mbox{\bf (a)}} &	\multicolumn{1}{l}{\mbox{\bf (b)}} \\ [-1cm]
\includegraphics[angle=0,width=3in]{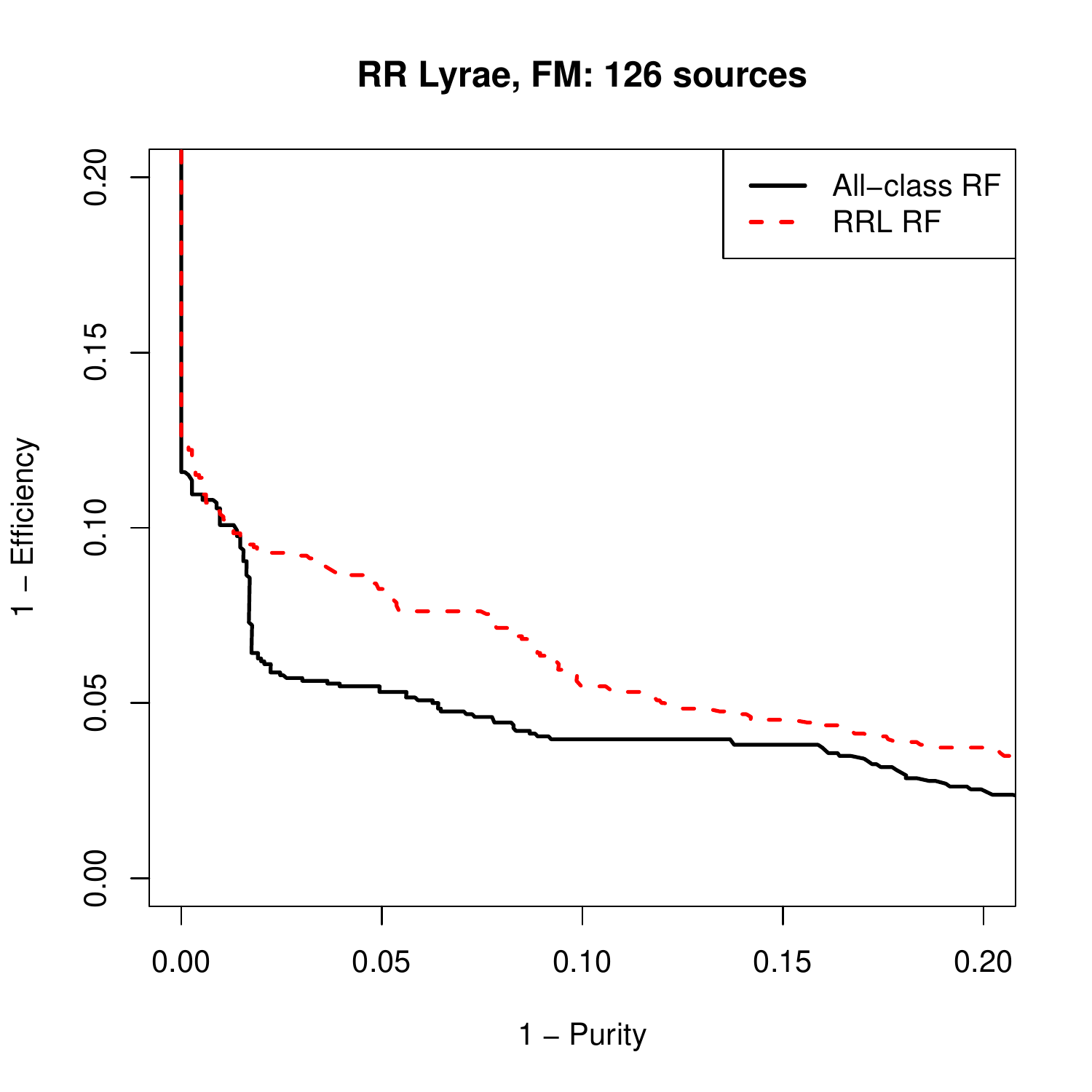}&
\includegraphics[angle=0,width=3in]{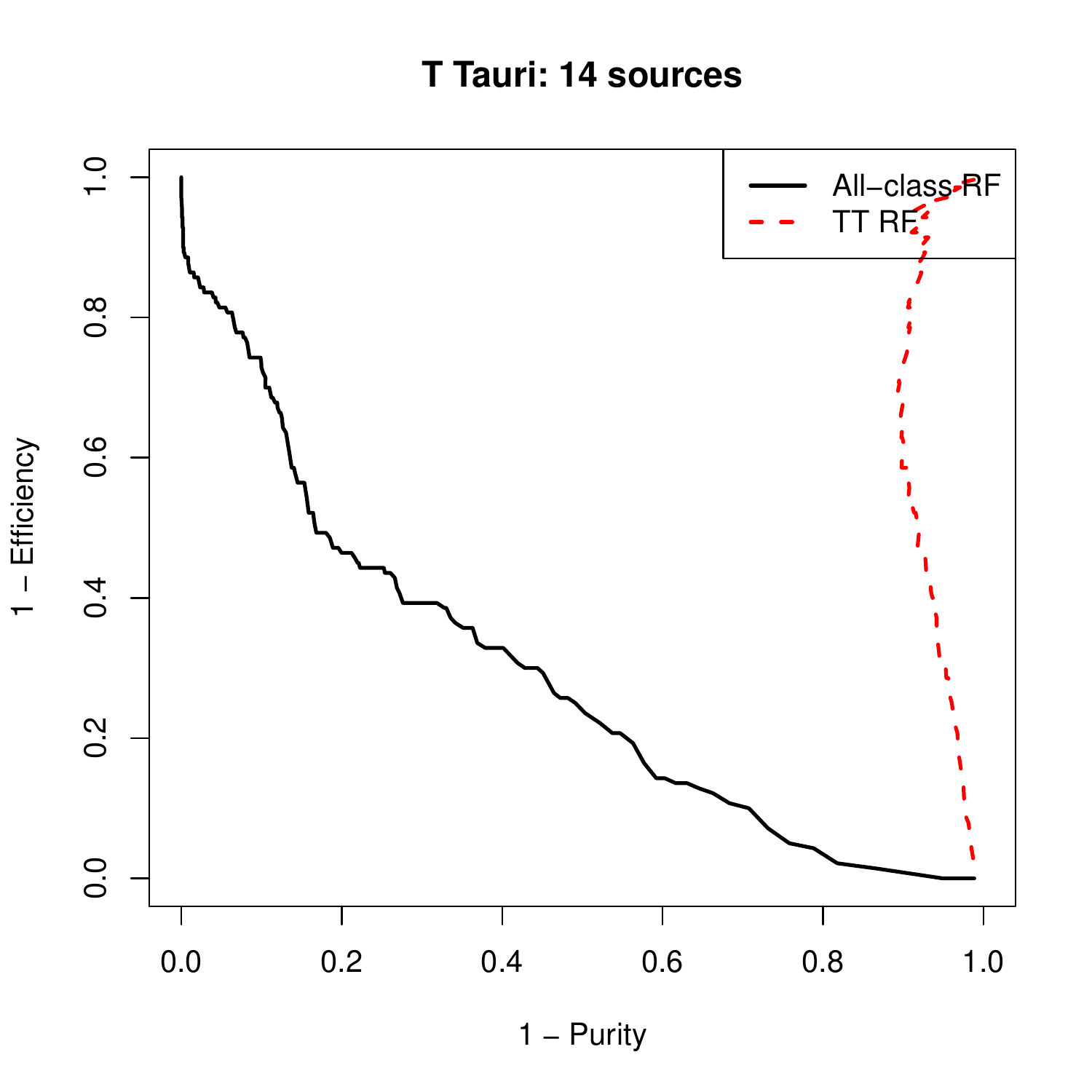}\\ 
\multicolumn{1}{l}{\mbox{\bf (c)}} &	\multicolumn{1}{l}{\mbox{\bf (d)}} \\ [-1cm]
\includegraphics[angle=0,width=3in]{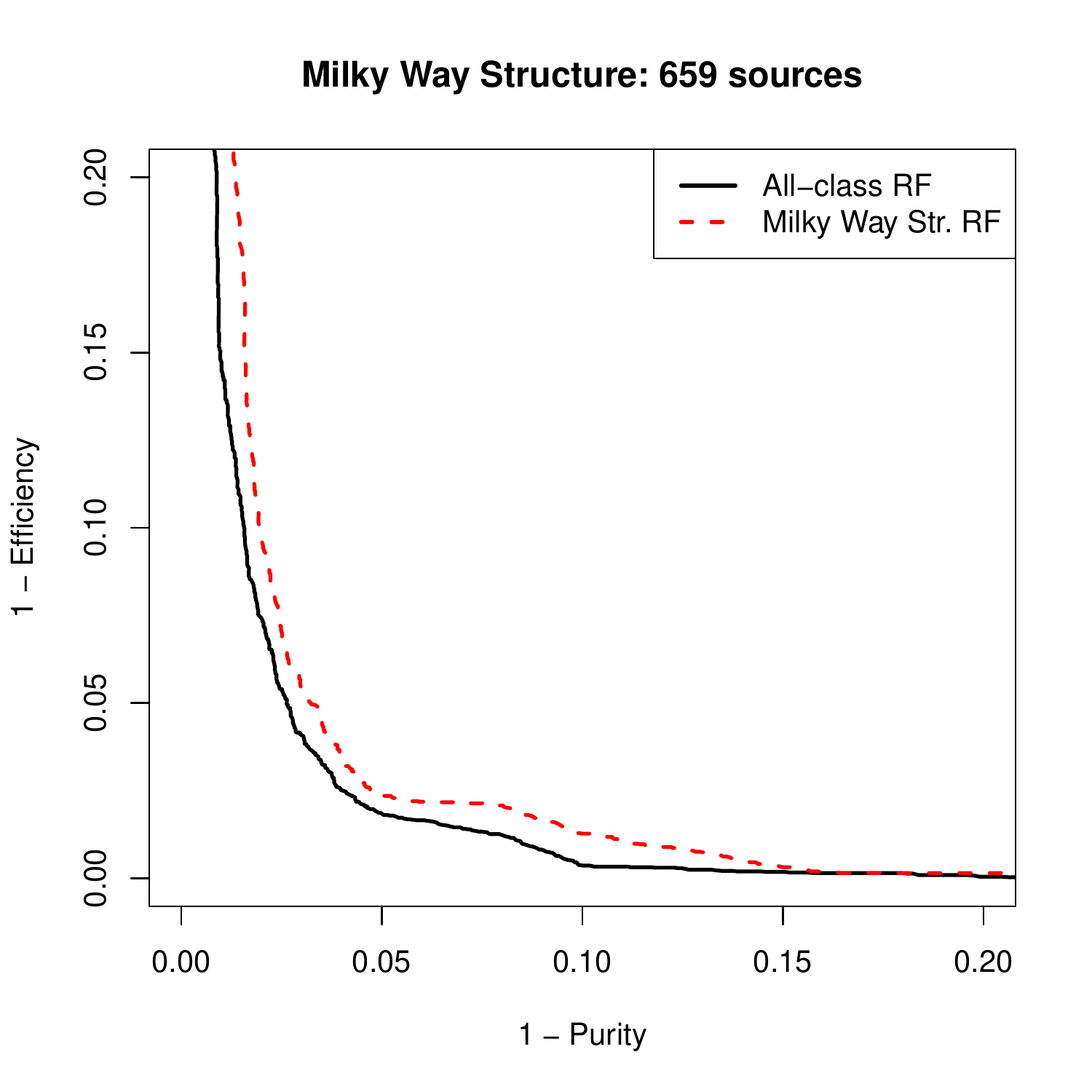}&
\includegraphics[angle=0,width=3in]{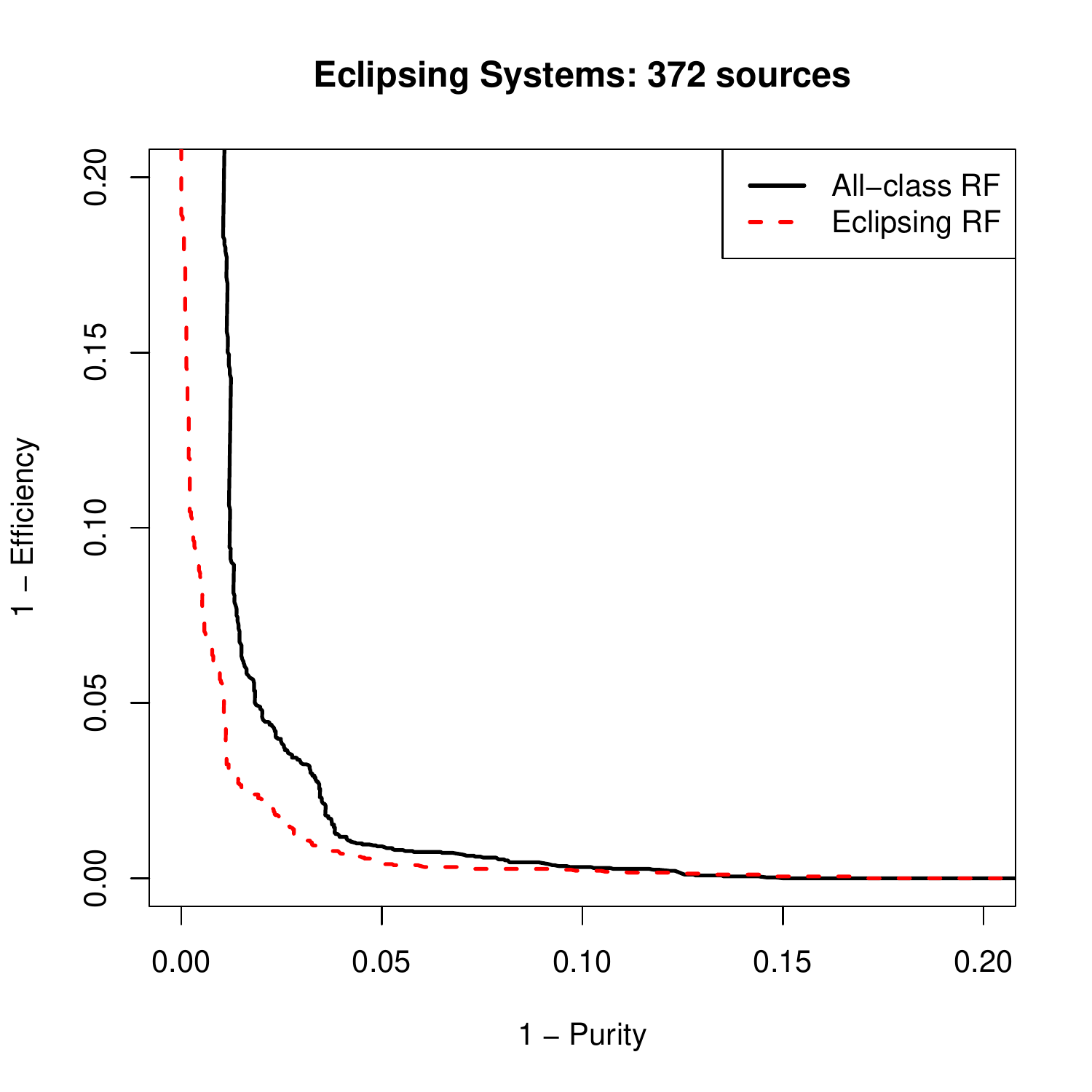}\\
\end{array}$
\end{center}
\caption{Cross-validated ROC curves, averaged over 10 cross validation trials, for four different science classes: (a) RR Lyrae, FM, (b) T Tauri, (c) Milky Way Structure, which includes all Mira, RR Lyrae, and Cepheid stars, and (d) Eclipsing Systems, which includes all Beta Persei, Beta Lyrae, and W Ursae Major stars.  Plotted is 1-Efficiency versus 1-Purity as a function of the class threshold.   The 25-class random forest ROC curve (solid black line) dominates the ROC curve of a one-versus-all random forest (dashed red line) for each science class except Eclipsing Systems.  For T Tauri, the all-class random forest is vastly superior to the T Tauri-specific classifier.   \label{fig:roc}}
\end{figure}

In figure \ref{fig:roc}, we plot the cross-validated ROC curve of the multi-class random forest for four different science classes: (a) RR Lyrae, FM, (b) T Tauri, (c) Milky Way Structure, which includes all Mira, RR Lyrae, and Cepheid stars, and (d) Eclipsing Systems, which includes all Beta Persei, Beta Lyrae, and W Ursae Major stars.  Each ROC curve shows the trade-off between the efficiency and purity of the samples.  At a 95\% purity, the estimated efficiency for RR Lyrae, FM is 94.7\%, for Milky Way Structure stars 98.2\%, and for Eclipsing Systems 99.1\%.  The T Tauri ROC curve is substantially worse than these other classes due to the small number of sources (note: inverse class-size weighting does not help in this problem because the ordering of the posterior class probabilities drives the ROC curve, not the magnitude of those probabilities).  Surprisingly, the 25-class random forest ROC curve dominates the ROC curve of a one-versus-all random forest for three of the four science classes, with vastly superior results for small classes.  This implies that our approach of predicting all class probabilities with one classifier is better than constructing a separate classifier for each science class of interest.

\subsection{Feature importance}

\begin{figure}
\begin{center}
\epsscale{.8}
\includegraphics[angle=0,scale=.7]{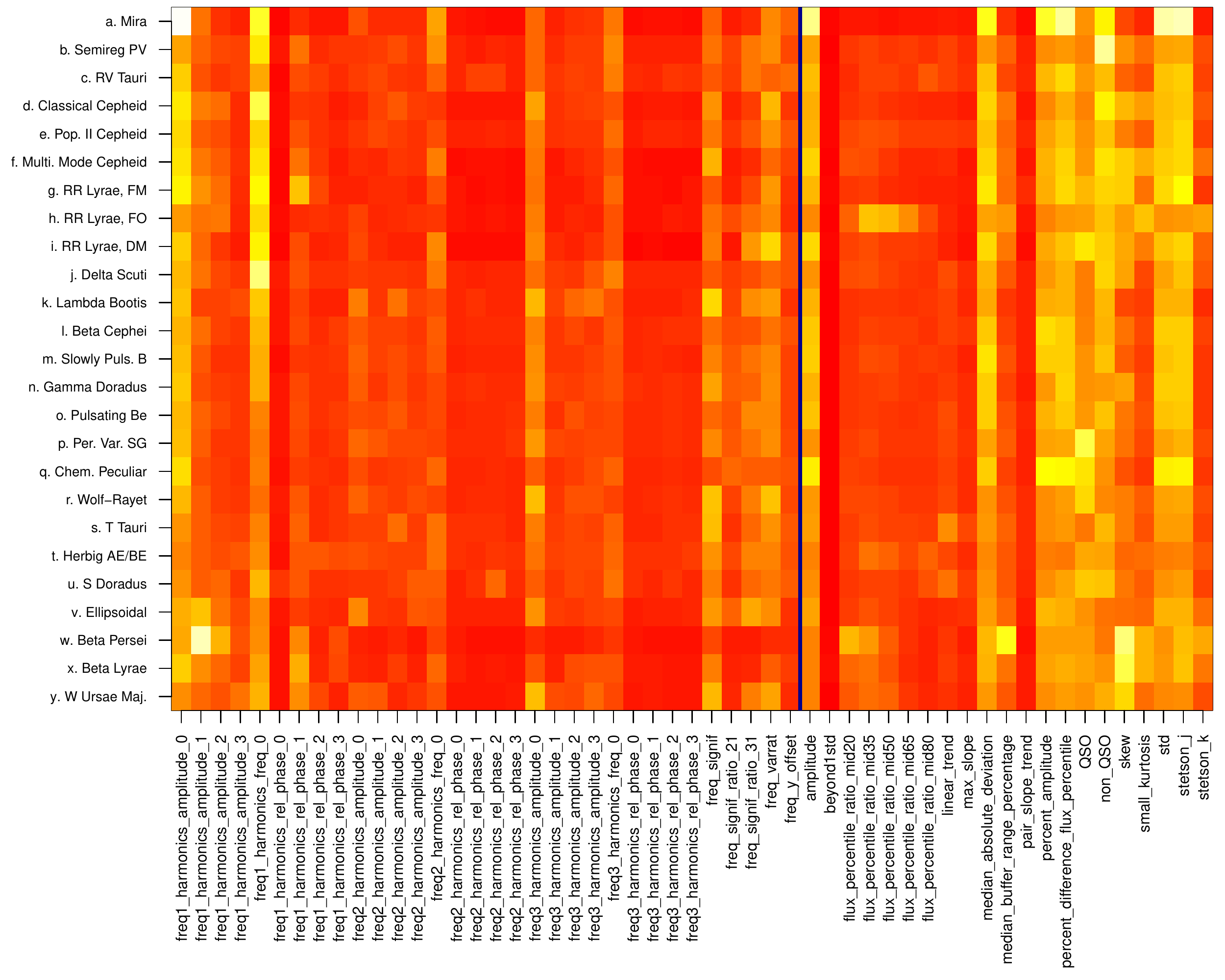}
\end{center}
\caption{Pairwise random forest feature importance.  Intensity is the square root of the proportion of instances of each class classified correctly because of that feature (compared to a replacement random permutation of that feature).  Features are split into periodic (left) and non-periodic (right) features.  The periodic features related to the first frequency are the most important, with higher-order frequencies and harmonics having smaller importance.  The non-periodic features related to spread and skew have high importance in the classifier, as do the QSO-variability features. \label{fig:featImp}}
\end{figure}

In \S\ref{ss:featimpmethod}, we described how to estimate the importance of each feature in a tree-based classifier.  In figure \ref{fig:featImp}, the importance of each feature from a pairwise random forest classifier is plotted, by class.  The intensity of each pixel depicts the proportion of instances of each class that are correctly-classified by using each particular feature in lieu of a feature containing random noise.  The intensities have been scaled to have the same mean across classes to mitigate the effects of unequal class sizes and are plotted on a square-root scale to decrease the influence of dominant features.  In independently comparing the performance of each feature to that of noise, the method does not account for correlations amongst features.  Consequentially, sets of features that measure similar properties--e.g. \verb median_absolute_deviation , \verb std , and \verb stetson_j  are all measures of the spread in the fluxes--may each have high importance, even though their combined importance is not much greater than each of their individual importances.

Figure \ref{fig:featImp} shows that a majority of the features used in this work have substantial importance in the random forest classifier for discerning at least one science class.  Close inspection of this figure illuminates the usefulness of certain features for distinguishing specific science classes.  As expected, the amplitude and frequency of the first harmonic, along with the features related to the spread and skew of the flux measurements, have the highest level of importance.  The flux amplitude is particularly important for classifying Mira stars, Cepheids, and RR Lyrae, which are distinguished by their large amplitudes.  The frequency of the first harmonic has high importance for most pulsating stars, likely because these classes have similar amplitudes but different frequencies.  The QSO variability feature is important for identifying eruptive sources, such as Periodically Variable Super Giants and Chemically Peculiar stars, because these stars generally have small values of the QSO feature compared to other variable star classes.   
 
 In addition, there are several features that have very low importance for distinguishing any of the science classes.  These features include the relative phase offsets for each of the harmonics, the amplitudes of the higher-order harmonics, and non-periodic features such as \verb beyond1std , \verb max_slope , and \verb pair_slope_trend .  We rerun the random forest classifier excluding the 14 features with the smallest feature importance (the excluded features are 9 relative phase offsets, \verb beyond1std , \verb max_slope , \verb pair_slope_trend , and the 65th and 80th middle flux percentiles.  Results show a cross-validated error rate of 22.8\% and a catastrophic error rate of 8.2\% (averaged over 10 repetitions of the random forest), which is consistent with the error rates of the random forest trained on all 53 features, showing the insensitivity of the random forest classifier to inclusion of features that carry little or no classification information.

\subsection{Classification of high-amplitude ($>$0.1 mag) sources}
\label{sec:highamp}

High-amplitude variable stars constitute many of the central scientific impetuses of current and future surveys. In particular, finding and classifying pulsational variables with known period-luminosity relationships (Mira, Cepheids and RR Lyrae) is a major thrust of the Large Synoptic Survey Telescope \citep{2009astro2010S.307W,2009arXiv0912.0201L}. 
Moreover, light curves from low-amplitude sources generally have a lower signal-to-noise ratio, making it more difficult to estimate their light-curve derived features and hence more difficult to obtain accurate classifications. 
Indeed, several of the classes in which we frequently make errors are populated by low-amplitude sources.

Here we classify only those light curves with amplitudes greater than 0.1 mag, removing low-amplitude classes from the sample. This results in the removal of 383 sources, or 25\% of the data.  After excluding these sources, we are left with a handful of classes with less than 7 sources.  Due to the difficulty in training (and cross-validating) a classifier for classes with such small amount of data, we ignore those classes, resulting in the exclusion of 19 more sources, bringing the total to 408 excluded sources, or 26\% of the entire data set.  We are left with 1134 sources in 16 science classes.

\begin{figure}
\begin{center}
\epsscale{.8}
\includegraphics[angle=0,scale=.6]{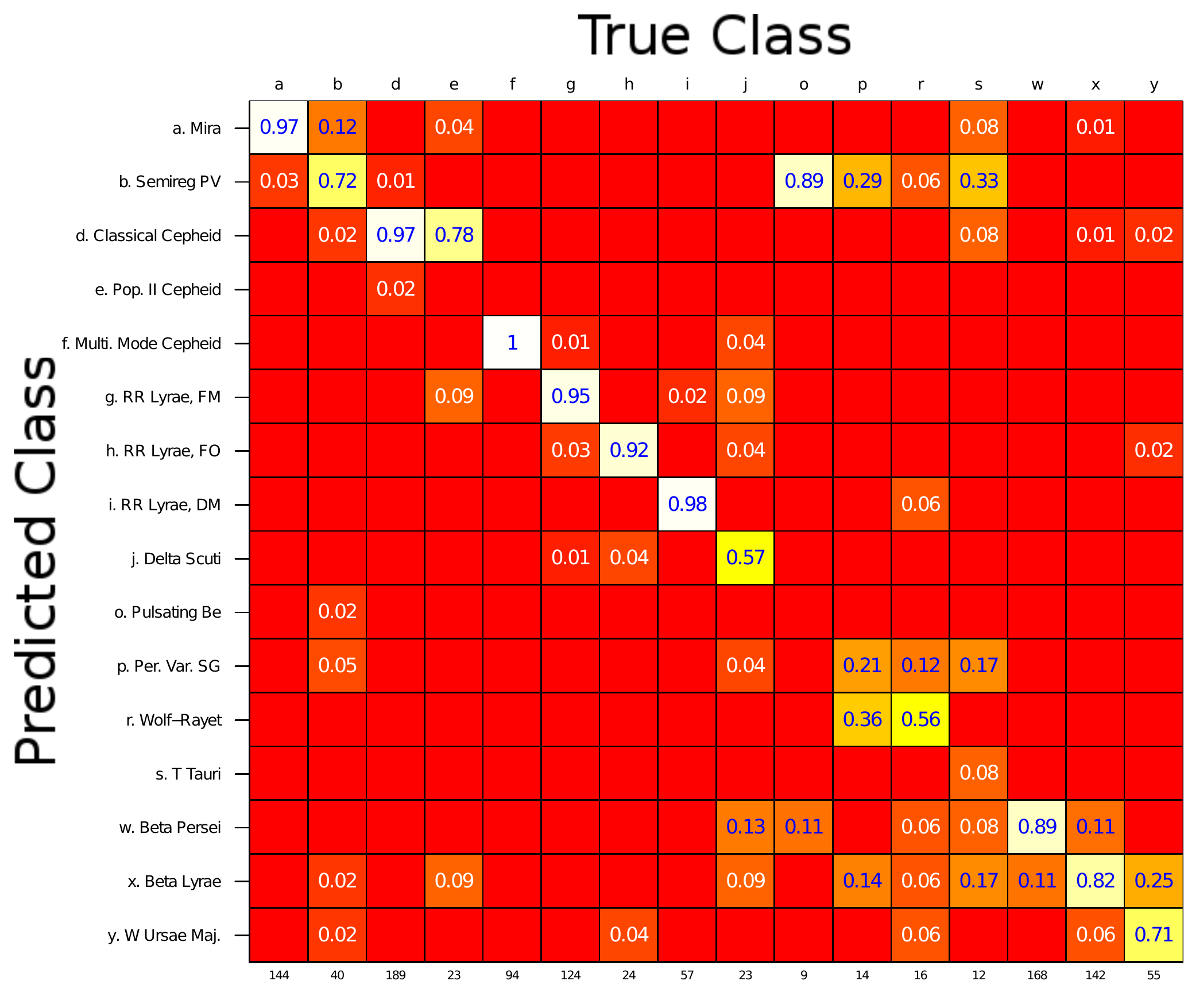}
\end{center}
\caption{Cross-validated confusion matrix for a random forest classifier applied only to the OGLE+Hipparcos sources with amplitude greater than 0.1 magnitude.  The overall error rate for this subset of data is 13.7\%, with catastrophic misclassification rate of 3.5\%. \label{fig:confmathighamp}}
\end{figure}

On this subset of data, our random forest classifier achieves a cross-validated misclassification rate of 13.7\%, a substantial improvement from the 22.8\% misclassification rate from the best classifier on the entire data set.  The catastrophic misclassification rate is only 3.5\%, compared to 8.0\% for the entire data set.  In figure \ref{fig:confmathighamp} the confusion matrix for the classifier is plotted.  The most prevalent error is misclassifying Pulsating Be and T Tauri stars as semi-regular pulsating variables.  On average 8 of the 9 Pulsating Be stars and 4 of the 12 T Tauri stars are misclassified as semi-regular pulsating variables.

\subsection{OGLE versus Hipparcos}
\label{ss:surveydependence}

Data from the sources that we classify in this section come from either the OGLE and Hipparcos surveys.  Specifically, there are 523 sources from five science classes whose data are from OGLE, while the data from the remaining 858  sources are from Hipparcos.  Our classifiers tend to perform much better for the OGLE data than for the Hipparcos sources: for the random forest classifier we obtain a 11.3\% error rate for OGLE data and 27.9\% error rate for Hipparcos.  

It is unclear whether the better performance for OGLE data is due to the relative ease at classifying the five science classes with OGLE data or because of differences in the survey specifications.  The sampling rate of the OGLE survey is three times higher than that of Hipparcos.   The average number of observations per OGLE source is 329, compared to 103 for Hipparcos, even though the average time baselines for the surveys are each near 1100 days.  OGLE observations have on average twice the flux as Hipparcos observations, but the flux measurement errors of OGLE light curves tend to be higher, making their respective signal-to-noise ratios similar (OGLE flux measurements have on average a SNR 1.25 times higher than Hipparcos SNRs).

To test whether the observed gains in accuracy between OGLE and Hipparcos sources is due to differences in the surveys or differences in the science classes observed by each survey, we run the following experiment:  For each OGLE light curve, we thin the flux measurements down to one-third of the original observations to mimic Hipparcos conditions and re-run the feature extraction pipeline and classifier using the thinned light curves.  Note that we do not add noise to the OGLE data because of the relative-similarity in average SNR between the surveys; the dominant difference between data of the two surveys is the sampling rate.  Results of the experiment are that the error rate for OGLE data increases to 13.0\%, an increase of only 1.7\% representing 9 more misclassified OGLE sources.  This value remains much lower than the Hipparcos error rate, showing that the better classifier performance for OGLE data is primarily driven by the ease of distinguishing those science classes.

\section{Conclusions and future work}

We have presented a thorough study of automated variable star classification from sparse and noisy single-band light curves.  In the 25-class problem considered by \citet{2007debo}, which includes all of the most important variable star science classes, we obtain a 24\% improvement over their best classifier in terms of misclassification error rate.   We attribute this improvement to all of the following advances:
\begin{itemize}
\item {\bf Better periodic feature estimation.}  Our Lomb-Scargle period-fitting code is both fast and accurate.  With the same random forest classifier, the average error rate using our periodic feature estimates is 23.8\%, compared to an error rate of 26.7\% using only Debosscher's period feature estimates, representing an improvement of 11\%.
\item {\bf Use of predictive non-periodic features.}  Simple summary statistics and more sophisticated model parameters give a significant improvement.  Using both our periodic and non-periodic features, the random forest error rate is 22.8\%, a 4\% improvement over using only our periodic features.
\item {\bf More accurate classification methods.}  All of the methods considered in this paper, save the single-tree models, achieve a statistically-significant improvement over Debosscher's best classifier.  Our random forest classifier, applied to the exact features used by that paper, achieves an 11\% improvement over their best error rate, 30\%.
\end{itemize}
Another contribution is our discussion of model validation through estimation of the expected prediction error on new data by cross-validation.  We presented cross-validation as a statistically-rigorous way to both tune a classifier and select between competing methods.  Indeed, all of the numbers quoted in this paper are 10-fold cross-validation estimates.

We have shown the adeptness of tree-based classifiers in the problem of variable star classification.  We demonstrated the superiority of the random forest classifier in terms of error rates, speed, and immunity to features with little useful classification information.  We outlined how to calculate the optimal probability threshold to obtain pure and complete samples of specified sub-classes, and showed that the multi-class random forest is often superior to the one-versus-all random forest in this problem.  We advocate the continued use of this method for other classification problems in astronomy.  

Furthermore, we described how the random forest classifier can be used to estimate the importance of each feature by computing the expected classification gains versus replacing that feature with random noise.  In the variable star classification problem, it was found that several non-periodic features have high importance.  A classifier built only on the non-periodic features still performs quite well, attaining 27.6\% error rate.

Finally, this paper is the first to use the known variable-star taxonomy both to train a classifier and evaluate its result.  We introduced two different classification methods to incorporate a hierarchical taxonomy: HSC, which builds a different classifier in each non-terminal node of the taxonomy, and HMC, which fits a single classifier, penalizing errors at smaller depths in the taxonomy more heavily.  We demonstrated that both of these methods perform well, in terms of classification rate and catastrophic error rate.  The class taxonomy was also used to construct the notion of catastrophic error rate, which considers as catastrophic any error made at the top level of the hierarchy.

Several open questions remain with regard to the automated classification of astronomical time series.  Many of these questions will be addressed by us in future publications, where we will expand on the methodology presented here and attempt to classify data from other surveys, such as the All Sky Automated Survey (ASAS), SDSS Stripe 82, and the Wide Angle Search for Planets (WASP).  Some of the questions that we will address are:
\begin{itemize}
	\item If we train a classifier on a set of objects from one (or multiple) survey(s), will that classifier be appropriate to predict the classes of objects from the new survey?  This question is of great importance because presumably a set of known (labeled) variable stars will be compiled from previous surveys to train a classifier for use on a new survey.
	\item What features are robust across a wide range of different surveys, each with different cadences?  If some sets of features are robust to survey design and cadence, those should be used in lieu of survey-dependent features in a classifier.  In this paper, we have excluded any feature that was blatantly survey-dependent (such as any that used mean flux), but this does not guarantee that some features will not have survey dependence.
    \item How does mis-labelled training data affect the classifier accuracy? Can mis-labelled data be effectively detected and cleaned from the classification set?  
	\item How can a classifier be trained to efficiently identify outliers/new types of variables?  Future surveys will unquestionably discover new science classes that do not fit under any of the training classes.  Recent methodology has been developed in the statistics literature for outlier discovery in random forest classifiers, and we plan to adapt this for variable star classification.
	\item How are the error rates of a classifier affected by computational limitations (where, perhaps some CPU-intensive or external server-dependent features are not used)?  In automated classification of astronomical sources, there is often a time sensitivity for follow-up observations.  Presumably there are more useful features for classification than the ones that we employed in this paper, but they may be expensive to compute or retrieve for each observation.  This trade-off between error rate and computation time must be explored.
\end{itemize}

Finally, as a longer-term goal, we are striving to develop methodology that can be used on a LSST-caliber survey.  This means that our methods must be fast enough to compute features and class probabilities for thousands of objects per night, work well at an LSST cadence, be applicable to multi-band light curves, and perform classification for all types of astronomical objects, including transients, variable stars, and QSOs.  Our task, looking forward, is to address each of these problems and develop methodology for fast and accurate classification for LSST.

\acknowledgements

The authors acknowledge the generous support of a CDI grant (\#0941742) from the National Science Foundation. J.S.B. and D.L.S.\ also thank the Las Cumbres Observatory for support during the early stages of this work.  We acknowledge helpful conversations with Dovi Poznanski, Tim Brown, John Rice, Noureddine El Kouri, Martin Wainwright, Elizabeth Purdom, Tamas Budavari, Kirk Borne, and Jeff Scargle. We thank J. Debosscher for assistance in obtaining the data from his paper.

\appendix

\section{Features estimated from light curves}
\label{app:features}

A description of the 32 periodic features computed using the methodology in \S \ref{sec:perfeat} is in table \ref{tab:LS}.
In addition to these 32 periodic features, we calculate 20 non-periodic
features for every light curve (\S \ref{sec:nonperfeat}).  These features are compiled in table \ref{tab:nonLS}.  These consist primarily of simple statistics that
can be calculated in the limit of few data points and also when no period is known in order to
characterize the flux variation distribution.  Where possible, we give the name of
the Python function that calculates the feature (e.g., skewness is from \verb scipy.stats.skew()  in
the Python SciPy module).

We begin with basic moment calculations using the observed photometric magnitude $mag$ vector for each source:

\begin{itemize}
\item \verb skew : skewness of the magnitudes: scipy.stats.skew();
\item \verb small_kurtosis : small sample kurtosis of the magnitudes\footnote{see \url{http://www.xycoon.com/peakedness\_small\_sample\_test\_1.htm}};
\item \verb std : standard deviation of the magnitudes: Numpy std();
\item \verb beyond1std :  the fraction ($\le 1$) of photometric magnitudes that lie above or below one std() from the weighted (by photometric errors) mean;
\item \verb stetson_j : \citet{1996stet} variability index, a robust standard deviation;
\item \verb stetson_k : \citet{1996stet} robust kurtosis measure.
\end{itemize}

We also calculate the following basic quantities using the magnitudes:

\begin{itemize}

\item \verb max_slope : examining successive (time-sorted) magnitudes, the maximal first difference (value of delta magnitude over delta time);
\item \verb amplitude : difference between the maximum and minimum magnitudes;
\item \verb median_absolute_deviation : ${\rm median}( | mag - {\rm median}(mag) | )$;
\item \verb median_buffer_range_percentage : fraction ($\le 1$) of photometric points within amplitude$/10$ of the median magnitude
\item \verb pair_slope_trend : considering the last 30 (time-sorted) measurements of source magnitude, the fraction of increasing first differences minus the fraction of decreasing first differences.
\end{itemize}

We also characterize the sorted flux $F=10^{-0.4 mag}$ distribution using percentiles, following \citet{2006eyer}.
If $F_{5,95}$ is the difference between 95\% and 5\% flux values, we calculate:

\begin{itemize}
\item \verb flux_percentile_ratio_mid20 : ratio $F_{40,60}/F_{5,95}$;
\item \verb flux_percentile_ratio_mid35 : ratio $F_{32.5,67.5}/F_{5,95}$;
\item \verb flux_percentile_ratio_mid50 : ratio $F_{25,75}/F_{5,95}$;
\item \verb flux_percentile_ratio_mid65 : ratio $F_{17.5,82.5}/F_{5,95}$;
\item \verb flux_percentile_ratio_mid80 : ratio $F_{10,90}/F_{5,95}$;
\item \verb percent_amplitude: the largest absolute departure from the median flux, divided by the median flux;
\item \verb percent_difference_flux_percentile: ratio of $F_{5,95}$ over the median flux.
\end{itemize}

Finally, useful for stochastically varying sources, we calculate the quasar similarity metrics from \citet{2010butl}:

\begin{itemize}
\item \verb QSO : quality of fit $\chi_{\rm QSO}^2/\nu$ for a quasar-like source, assuming $mag=19$;
\item \verb non_QSO : quality of fit for a non-quasar-like source (related to NULL value of QSO).
\end{itemize}

\begin{deluxetable}{lc} 
\tablecolumns{2} 
\tablewidth{0pc} 
\tablecaption{Periodic features extracted from light curves using generalized Lomb-Scargle} 
\tablehead{ 
\colhead{Feature}    &  \colhead{Description\tablenotemark{a} }   }
\startdata 
\verb freq1_harmonics_amplitude_0 & $A_{1,1}$\tablenotemark{b} \\
\verb freq1_harmonics_amplitude_1 & $A_{1,2}$   \\
\verb freq1_harmonics_amplitude_2 &   $A_{1,3}$ \\
\verb freq1_harmonics_amplitude_3 &  $A_{1,4}$  \\
\verb freq1_harmonics_freq_0 &   $f_{1}$\tablenotemark{c}\\
\verb freq1_harmonics_rel_phase_0 &  $PH_{1,1}$\tablenotemark{d}\\
\verb freq1_harmonics_rel_phase_1 &  $PH_{1,2}$ \\
\verb freq1_harmonics_rel_phase_2 &   $PH_{1,3}$\\
\verb freq1_harmonics_rel_phase_3 &   $PH_{1,4}$\\
\verb freq2_harmonics_amplitude_0 &   $A_{2,1}$\\
\verb freq2_harmonics_amplitude_1 &   $A_{2,2}$ \\
\verb freq2_harmonics_amplitude_2 &    $A_{2,3}$\\
\verb freq2_harmonics_amplitude_3 &    $A_{2,4}$\\
\verb freq2_harmonics_freq_0 &  $f_{2}$  \\
\verb freq2_harmonics_rel_phase_0 &  $PH_{2,1}$  \\
\verb freq2_harmonics_rel_phase_1 &   $PH_{2,2}$ \\
\verb freq2_harmonics_rel_phase_2 &   $PH_{2,3}$ \\
\verb freq2_harmonics_rel_phase_3 &   $PH_{2,4}$ \\
\verb freq3_harmonics_amplitude_0 &    $A_{3,1}$\\
\verb freq3_harmonics_amplitude_1 &    $A_{3,2}$\\
\verb freq3_harmonics_amplitude_2 &    $A_{3,3}$\\
\verb freq3_harmonics_amplitude_3 &    $A_{3,4}$\\
\verb freq3_harmonics_freq_0 &   $f_{3}$ \\
\verb freq3_harmonics_rel_phase_0 &   $PH_{3,1}$ \\
\verb freq3_harmonics_rel_phase_1 &   $PH_{3,2}$ \\
\verb freq3_harmonics_rel_phase_2 &   $PH_{3,3}$ \\
\verb freq3_harmonics_rel_phase_3 &   $PH_{3,4}$ \\
\verb freq_signif &  Significance of $f_1$ vs. null hypothesis of white noise with no periodic \\
& variation, computed using a Student's-$T$ distribution \\
\verb freq_signif_ratio_21 & Ratio of significance of $f_2$ vs.  null to  $f_1$ vs. null   \\
\verb freq_signif_ratio_31 &  Ratio of significance of $f_3$ vs. null to  $f_1$ vs. null  \\
\verb freq_varrat &  Ratio of the variance after, to the variance before subtraction \\
 &    of the fit with $f_1$ and its 4 harmonics \\
\verb freq_y_offset & $c$  
\enddata
\tablenotetext{a}{Notation from discussion of Lomb-Scargle periodic feature extraction in \S \ref{sec:perfeat} is used.}
\tablenotetext{b}{All amplitudes are in units of magnitude.}
\tablenotetext{c}{All frequencies are in units of cycles/day.}
\tablenotetext{d}{All relative phases are unitless ratios.}
\label{tab:LS}
\end{deluxetable}%

\begin{deluxetable}{ll} 
\tablecolumns{2} 
\tablewidth{0pc} 
\tablecaption{Non-periodic features extracted from light curves} 
\tablehead{ 
\colhead{Feature}    &  \colhead{Description}   }
\startdata 
\verb amplitude      &  Half the difference between the maximum and the minimum magnitude\\
\verb beyond1std    & Percentage of points beyond one st. dev. from the weighted mean \\
\verb flux_percentile_ratio_mid20	    & Ratio of flux percentiles (60th - 40th) over (95th - 5th)\\
\verb flux_percentile_ratio_mid35	     & Ratio of flux percentiles (67.5th - 32.5th) over (95th - 5th)\\
\verb flux_percentile_ratio_mid50	    & Ratio of flux percentiles (75th - 25th) over (95th - 5th)\\
\verb flux_percentile_ratio_mid65	    & Ratio of flux percentiles (82.5th - 17.5th) over (95th - 5th)\\
\verb flux_percentile_ratio_mid80	    & Ratio of flux percentiles (90th - 10th) over (95th - 5th)\\
\verb linear_trend	    & Slope of a linear fit to the light curve fluxes\\
\verb max_slope	    & Maximum absolute flux slope between two consecutive observations\\
\verb median_absolute_deviation	    &   Median discrepancy of the fluxes from the median flux \\
\verb median_buffer_range_percentage     & Percentage of fluxes within 20\% of the amplitude from the median\\
\verb pair_slope_trend	    & Percentage of all pairs of consecutive flux measurements that have positive slope \\
\verb percent_amplitude	    & Largest percentage difference between either the max or min magnitude and the median\\
\verb percent_difference_flux_percentile	    & Diff. between the 2nd \& 98th flux percentiles, converted to magnitude\tablenotemark{a} \\
\verb QSO & Quasar variability metric in \citet{2010butl} \\
\verb non_QSO & Non-quasar variability metric in \citet{2010butl} \\
\verb skew		    & Skew of the fluxes\\
\verb small_kurtosis		    &  Kurtosis of the fluxes, reliable down to a small number of epochs\\
\verb std		    & Standard deviation of the fluxes \\
\verb stetson_j	    & Welch-Stetson variability index J\tablenotemark{b}\\
\verb stetson_k	    & Welch-Stetson variability index K\tablenotemark{b}
\enddata
\tablenotetext{a}{\citet{2005eyer}}
\tablenotetext{b}{\citet{1996stet}}
\label{tab:nonLS}
\end{deluxetable}%

\section{Doubly-labeled stars}
\label{app:double}

As discussed in \S\ref{sec:data}, there are 25 sources with more than one class label. Five of these objects are labeled as both S Doradus and periodically-variable super giants.  We assign these to the S Doradus class because S Doradus is a subclass of periodically-variable super giants.   Two other sources, V* BF Ori and HD 97048 were verified to be Herbig AE/BE type stars by \citet{2010grin} and \citet{2007doer}, respectively.  RY Lep is listed incorrectly in Simbad as an Agol-type eclipsing system (its original 50-year old classification) and as RR Lyrae/Delta Scuti. \citet{2009MNRAS.394..995D} clearly establish this source as a high-ampltiude $\delta$-Scuti in a binary system. Likewise, HIP 99252 is a $\delta$-Scuti star \citep{2000A&AS..144..469R} and not also an RR Lyrae. HIP 30326 (=V Mon) is a Mira variable and not also part of the RV Tauri class \citep{2008MNRAS.386..313W}. HD 210111 (listed also as delta Scuti) is a $\lambda$ Bootis star \citep{2008CoAst.153...49P}.  HD 165763 is a WR star that is not periodic to a high-degree of confidence \citep{2008ApJ...679L..45M}. Likewise, WR40 is not periodic as it was once believed \citep{1994AJ....108..678M}. We found no reference to the periodic nature of the WR star HD 156385. The remaining 11 sources were of ambiguous class or truly deserved of two classes, and were excluded from the training and testing sample.   These stars are listed in table \ref{tab:throwouts}, and briefly mentioned below:

\begin{itemize}
	\item {\it HD 136488 = HIP 75377}: This is a WR with a measured periodicity in Hipparcos data \citep{2002MNRAS.331...45K}.
		
	\item {\it HD 94910 = AG Car}: This is a WR is thought to be a ``hot, quiescent state LBV'' \citep{2010A&A...514A..87C}.
	
	\item {\it HD 50896 = EZ CMa = WR 6}: Listed as both a WR and periodic variable super giant, this is a prototype WR has a known periodic variability possibly due to a binary companion \citep{2007AJ....133.2859F}.
	
	\item {\it HD 37151}: Slowly pulsating B-stars (SPBs) occur in the same instability strip of the H-R diagram as chemically peculiar B stars (with variability associated with rotation and magnetic fields). The classification of this star is particularly ambiguous (see \citealt{2007A&A...466..269B} and references therein).

    \item {\it HD 35715}  Is a cataloged Beta Cepheid but is also an ellipsoidal variable (\citealt{2005stan}). Pulsation was not detected photometrically but in line profiles.

    \item {\it HIP34042 (Z CMa A)} Is a binary system consisting of a Herbig component (A) and an FU Orionis star (B) in a close orbit.  Determining the photometric contributions of the two components separately is problematic because of strong variability and the small angular separation (\citealt{2002mill}).

    \item {\it TV Lyn, UY Cam, V1719 Cyg (HD 200925) \& V753 Cen} Ambiguous classification (within the literature) of RR Lyrae or Delta Scuti type stars.

    \item {\it IK Hya} Is a short-period Pop. II Cepheid, making it likely a BL Herculis type star.
\end{itemize}

We trained a random forest classifier on the features of only the single-labeled stars, and then predicted the class probabilities of each label for the doubly-labeled stars.  This experiment shows that for 7 of the 11 sources, our classifier predicts one of the two classes that the star was originally labeled as.
Note that recently, several methods have been developed to classify data with multiple labels (multi-label classification, see \citealt{2007tsou}), but for the purposes of this work we choose to only perform single-label classification, and remove these data from our sample to avoid biases from inaccurate training labels.

\begin{deluxetable}{lllllll} 
\tablecolumns{6} 
\tablewidth{0pc} 
\tablecaption{Doubly-labelled sources excluded from the sample for both training and testing purposes.} 
\tablehead{ 
\colhead{RA} & \colhead{Dec} & \colhead{ID} & \colhead{Class 1 \tablenotemark{a}} & \colhead{Class 2 \tablenotemark{a}} &  \colhead{RF best class \tablenotemark{b}}& \colhead{RF 2nd best class \tablenotemark{c}}}
\startdata 
05 26 50.2284 & +03 05 44.428& HD 35715   &  l. Beta Cephei  & v. Ellipsoidal &m. Slowly Puls. B       (0.214)&       v. Ellipsoidal        (0.164) \\
05 36 06.2333 & $-$07 23 47.320 &HD 37151  &   q. Chem. Peculiar & m. Slowly Puls. B&  p. Per. Var. SG       (0.211)&     n. Gamma Doradus        (0.155)\\
06 54 13.0441 & $-$23 55 42.011 & HD 50896 & r. Wolf-Rayet & p. Per. Var. SG & v. Ellipsoidal       (0.181)&      o. Pulsating Be        (0.145)\\
07 03 43.1619 &$-$11 33 06.209  & HIP 34042   &  t. Herbig AE/BE  &     s. T Tauri     & b. Semireg PV       (0.287)& d. Classical Cepheid        (0.114)\\
07 33 31.7288 &+47 48 09.823 & TV Lyn & j. Delta Scuti & h. RR Lyrae, FO & h. RR Lyrae, FO       (0.686) &      g. RR Lyrae, FM         (0.12)\\
07 58 58.8801 &+72 47 15.411 & UY Cam &   j. Delta Scuti & h. RR Lyrae, FO    & h. RR Lyrae, FO       (0.236)&       j. Delta Scuti        (0.222)\\
10 56 11.5763 & $-$60 27 12.815 & HD 94910         & r. Wolf-Rayet & u. S Doradus & b. Semireg PV       (0.258)&              a. Mira        (0.243)\\
11 51 15.3088 & $-$55 48 15.795  & V753 Cen  &  j. Delta Scuti  &h. RR Lyrae, FO   & h. RR Lyrae, FO       (0.469)&       j. Delta Scuti        (0.151)\\
12 04 47.2738 &$-$27 40 43.295 & IK Hya  &e. Pop. II Cepheid & g. RR Lyrae, FM&g. RR Lyrae, FM       (0.456)&      y. W Ursae Maj.          (0.100)\\
15 24 11.3087 & $-$62 40 37.567 & HD 136488             & p. Per. Var. SG & r. Wolf-Rayet &    r. Wolf-Rayet       (0.161) &     p. Per. Var. SG        (0.147)\\
21 04 32.9211 & +50 47 03.276  & V1719 Cyg    &  j. Delta Scuti  &h. RR Lyrae, FO    & j. Delta Scuti       (0.315)   &   y. W Ursae Maj.        (0.226)
 \enddata
\tablenotetext{a}{Classes determined by \citet{2007debo}.}
\tablenotetext{b}{Most likely class determined by a random forest classifier.  Posterior probability in parentheses. }
\tablenotetext{c}{Second-best candidate class determined by a random forest classifier.  Posterior probability in parentheses. }
 \label{tab:throwouts}
 \end{deluxetable}

\bibliography{TSclassify}

\begin{thebibliography}{69}
\expandafter\ifx\csname natexlab\endcsname\relax\def\natexlab#1{#1}\fi

\bibitem[{{Bailey} {et~al.}(2007){Bailey}, {Aragon}, {Romano}, {Thomas},
  {Weaver}, \& {Wong}}]{2007bail}
{Bailey}, S., {Aragon}, C., {Romano}, R., {Thomas}, R.~C., {Weaver}, B.~A., \&
  {Wong}, D. 2007, \apj, 665, 1246

\bibitem[{{Ball} {et~al.}(2006){Ball}, {Brunner}, {Myers}, \&
  {Tcheng}}]{2006ball}
{Ball}, N.~M., {Brunner}, R.~J., {Myers}, A.~D., \& {Tcheng}, D. 2006, \apj,
  650, 497

\bibitem[{{Barning}(1963)}]{1963barn}
{Barning}, F.~J.~M. 1963, \bain, 17, 22

\bibitem[{Blockeel {et~al.}(2006)Blockeel, Schietgat, Struyf, D{\v{z}}eroski,
  \& Clare}]{2006bloc}
Blockeel, H., Schietgat, L., Struyf, J., D{\v{z}}eroski, S., \& Clare, A. 2006,
  Knowledge Discovery in Databases: PKDD 2006, 18

\bibitem[{{Blomme} {et~al.}(2010){Blomme}, {Debosscher}, {De Ridder}, {Aerts},
  {Gilliland}, {Christensen-Dalsgaard}, {Kjeldsen}, {Brown}, {Borucki}, {Koch},
  {Jenkins}, {Kurtz}, {Stello}, {Stevens}, {Suran}, \& {Derekas}}]{2010blom}
{Blomme}, J., {et~al.} 2010, \apjl, 713, L204

\bibitem[{Breiman(1996)}]{1996brei}
Breiman, L. 1996, Machine learning, 24, 123

\bibitem[{Breiman(2001)}]{2001brei}
---. 2001, Machine learning, 45, 5

\bibitem[{Breiman {et~al.}(1984)Breiman, Friedman, Olshen, \& Stone}]{1984brei}
Breiman, L., Friedman, J., Olshen, R., \& Stone, C. 1984, Wadsworth Inc, 67

\bibitem[{{Brett} {et~al.}(2004){Brett}, {West}, \& {Wheatley}}]{2004bret}
{Brett}, D.~R., {West}, R.~G., \& {Wheatley}, P.~J. 2004, \mnras, 353, 369

\bibitem[{{Briquet} {et~al.}(2007){Briquet}, {Hubrig}, {De Cat}, {Aerts},
  {North}, \& {Sch{\"o}ller}}]{2007A&A...466..269B}
{Briquet}, M., {Hubrig}, S., {De Cat}, P., {Aerts}, C., {North}, P., \&
  {Sch{\"o}ller}, M. 2007, \aap, 466, 269

\bibitem[{Burman(1989)}]{1989burm}
Burman, P. 1989, Biometrika, 76, 503

\bibitem[{{Butler} \& {Bloom}(2010)}]{2010butl}
{Butler}, N.~R., \& {Bloom}, J.~S. 2010, arXiv e-print 1008.3143

\bibitem[{Cesa-Bianchi {et~al.}(2006)Cesa-Bianchi, Gentile, \&
  Zaniboni}]{2006cesa}
Cesa-Bianchi, N., Gentile, C., \& Zaniboni, L. 2006, The Journal of Machine
  Learning Research, 7, 31

\bibitem[{Cheeseman \& Stutz(1996)}]{1996chee}
Cheeseman, P., \& Stutz, J. 1996, Advances in knowledge discovery and data
  mining, 180

\bibitem[{{Clark} {et~al.}(2010){Clark}, {Ritchie}, \&
  {Negueruela}}]{2010A&A...514A..87C}
{Clark}, J.~S., {Ritchie}, B.~W., \& {Negueruela}, I. 2010, \aap, 514, A87

\bibitem[{{Covey} {et~al.}(2007){Covey}, {Ivezi{\'c}}, {Schlegel},
  {Finkbeiner}, {Padmanabhan}, {Lupton}, {Ag{\"u}eros}, {Bochanski}, {Hawley},
  {West}, {Seth}, {Kimball}, {Gogarten}, {Claire}, {Haggard}, {Kaib},
  {Schneider}, \& {Sesar}}]{2007AJ....134.2398C}
{Covey}, K.~R., {et~al.} 2007, \aj, 134, 2398

\bibitem[{{Debosscher} {et~al.}(2007){Debosscher}, {Sarro}, {Aerts}, {Cuypers},
  {Vandenbussche}, {Garrido}, \& {Solano}}]{2007debo}
{Debosscher}, J., {Sarro}, L.~M., {Aerts}, C., {Cuypers}, J., {Vandenbussche},
  B., {Garrido}, R., \& {Solano}, E. 2007, \aap, 475, 1159

\bibitem[{{Derekas} {et~al.}(2009){Derekas}, {Kiss}, {Bedding}, {Ashley},
  {Cs{\'a}k}, {Danos}, {Fernandez}, {F{\H u}r{\'e}sz}, {M{\'e}sz{\'a}ros},
  {Szab{\'o}}, {Szak{\'a}ts}, {Sz{\'e}kely}, \&
  {Szatm{\'a}ry}}]{2009MNRAS.394..995D}
{Derekas}, A., {et~al.} 2009, \mnras, 394, 995

\bibitem[{{Doering} {et~al.}(2007){Doering}, {Meixner}, {Holfeltz}, {Krist},
  {Ardila}, {Kamp}, {Clampin}, \& {Lubow}}]{2007doer}
{Doering}, R.~L., {Meixner}, M., {Holfeltz}, S.~T., {Krist}, J.~E., {Ardila},
  D.~R., {Kamp}, I., {Clampin}, M.~C., \& {Lubow}, S.~H. 2007, \aj, 133, 2122

\bibitem[{{Eads} {et~al.}(2004){Eads}, {Williams}, {Theiler}, {Porter},
  {Harvey}, {Perkins}, {Brumby}, \& {David}}]{2004SPIE.5200...79E}
{Eads}, D.~R., {Williams}, S.~J., {Theiler}, J., {Porter}, R., {Harvey}, N.~R.,
  {Perkins}, S.~J., {Brumby}, S.~P., \& {David}, N.~A. 2004, in Society of
  Photo-Optical Instrumentation Engineers (SPIE) Conference Series, ed.
  {B.~Bosacchi, D.~B.~Fogel, and J.~C.~Bezdek}, Vol. 5200, 79--90

\bibitem[{{Eyer}(2005)}]{2005eyer}
{Eyer}, L. 2005, in ESA Special Publication, Vol. 576, The Three-Dimensional
  Universe with Gaia, ed. {C.~Turon, K.~S.~O'Flaherty, and M.~A.~C.~Perryman},
  513

\bibitem[{{Eyer}(2006)}]{2006eyer}
{Eyer}, L. 2006, in Astronomical Society of the Pacific Conference Series, Vol.
  349, Astrophysics of Variable Stars, ed. {C.~Aerts and C.~Sterken}, 15

\bibitem[{{Eyer} \& {Blake}(2005)}]{2005eyer1}
{Eyer}, L., \& {Blake}, C. 2005, \mnras, 358, 30

\bibitem[{{Eyer} \& {Mowlavi}(2008)}]{em08}
{Eyer}, L., \& {Mowlavi}, N. 2008, Journal of Physics Conference Series, 118,
  012010

\bibitem[{{Eyer} {et~al.}(2008){Eyer}, {Jan}, {Dubath}, {Nienartovicz},
  {Blomme}, {Debosscher}, {De Ridder}, {Lopez}, \&
  {Sarro}}]{2008AIPC.1082..257E}
{Eyer}, L., {et~al.} 2008, in American Institute of Physics Conference Series,
  Vol. 1082, American Institute of Physics Conference Series, ed.
  {C.~A.~L.~Bailer-Jones}, 257--262

\bibitem[{{Flores} {et~al.}(2007){Flores}, {Koenigsberger}, {Cardona}, \& {de
  la Cruz}}]{2007AJ....133.2859F}
{Flores}, A., {Koenigsberger}, G., {Cardona}, O., \& {de la Cruz}, L. 2007,
  \aj, 133, 2859

\bibitem[{Freund \& Schapire(1996)}]{1996freu}
Freund, Y., \& Schapire, R. 1996, in International Conference on Machine
  Learning, 148--156

\bibitem[{Friedman(1996)}]{1996frie}
Friedman, J. 1996, technical report, Department of Statistics, Stanford
  University

\bibitem[{Friedman(2001)}]{2001frie}
Friedman, J.~H. 2001, The Annals of Statistics, 29, pp. 1189

\bibitem[{{Gregory}(2005)}]{gregory05}
{Gregory}, P.~C. 2005, {Bayesian Logical Data Analysis for the Physical
  Sciences: A Comparative Approach with `Mathematica' Support} (Cambridge
  University Press)

\bibitem[{{Grinin} {et~al.}(2010){Grinin}, {Rostopchina}, {Barsunova}, \&
  {Demidova}}]{2010grin}
{Grinin}, V.~P., {Rostopchina}, A.~N., {Barsunova}, O.~Y., \& {Demidova}, T.~V.
  2010, Astrophysics, 53, 367

\bibitem[{Hastie \& Tibshirani(1998)}]{1998hast}
Hastie, T., \& Tibshirani, R. 1998, Annals of Statistics, 26, 451

\bibitem[{Hastie {et~al.}(2009)Hastie, Tibshirani, \& Friedman}]{2009hast}
Hastie, T., Tibshirani, R., \& Friedman, J. 2009, The Elements of Statistical
  Learning: Data Mining, Inference, and Prediction., 2nd edn. (Springer, New
  York)

\bibitem[{{Horne} \& {Baliunas}(1986)}]{hb86}
{Horne}, J.~H., \& {Baliunas}, S.~L. 1986, \apj, 302, 757

\bibitem[{{Ivezi{\'c}} {et~al.}(2007){Ivezi{\'c}}, {Smith}, {Miknaitis}, {Lin},
  {Tucker}, {Lupton}, {Gunn}, {Knapp}, {Strauss}, {Sesar}, {Doi}, {Tanaka},
  {Fukugita}, {Holtzman}, {Kent}, {Yanny}, {Schlegel}, {Finkbeiner},
  {Padmanabhan}, {Rockosi}, {Juri{\'c}}, {Bond}, {Lee}, {Stoughton}, {Jester},
  {Harris}, {Harding}, {Morrison}, {Brinkmann}, {Schneider}, \&
  {York}}]{2007AJ....134..973I}
{Ivezi{\'c}}, {\v Z}., {et~al.} 2007, \aj, 134, 973

\bibitem[{{Kessler} {et~al.}(2010){Kessler}, {Bassett}, {Belov}, {Bhatnagar},
  {Campbell}, {Conley}, {Frieman}, {Glazov}, {Gonz{\'a}lez-Gait{\'a}n},
  {Hlozek}, {Jha}, {Kuhlmann}, {Kunz}, {Lampeitl}, {Mahabal}, {Newling},
  {Nichol}, {Parkinson}, {Philip}, {Poznanski}, {Richards}, {Rodney}, {Sako},
  {Schneider}, {Smith}, {Stritzinger}, \& {Varughese}}]{2010kess}
{Kessler}, R., {et~al.} 2010, \pasp, 122, 1415

\bibitem[{Knerr {et~al.}(1990)Knerr, Personnaz, Dreyfus, Fogelman, Agresti,
  Ajiz, Jennings, Alizadeh, Alizadeh, \& Haeberly}]{1990kner}
Knerr, S., {et~al.} 1990, Optimization Methods and Software, 1, 23

\bibitem[{{Koen} \& {Eyer}(2002)}]{2002MNRAS.331...45K}
{Koen}, C., \& {Eyer}, L. 2002, \mnras, 331, 45

\bibitem[{{Lomb}(1976)}]{1976lomb}
{Lomb}, N.~R. 1976, \apss, 39, 447

\bibitem[{{LSST Science Collaborations} {et~al.}(2009){LSST Science
  Collaborations}, {Abell}, {Allison}, {Anderson}, {Andrew}, {Angel}, {Armus},
  {Arnett}, {Asztalos}, {Axelrod}, \& et~al.}]{2009arXiv0912.0201L}
{LSST Science Collaborations} {et~al.} 2009, arXiv e-print 0912.0201

\bibitem[{{Mahabal} {et~al.}(2008){Mahabal}, {Djorgovski}, {Turmon}, {Jewell},
  {Williams}, {Drake}, {Graham}, {Donalek}, {Glikman}, \& {Palomar-QUEST
  team}}]{2008AN....329..288M}
{Mahabal}, A., {et~al.} 2008, Astronomische Nachrichten, 329, 288

\bibitem[{{Marchenko} {et~al.}(1994){Marchenko}, {Antokhin}, {Bertrand},
  {Lamontagne}, {Moffat}, {Piceno}, \& {Matthews}}]{1994AJ....108..678M}
{Marchenko}, S.~V., {Antokhin}, I.~I., {Bertrand}, J., {Lamontagne}, R.,
  {Moffat}, A.~F.~J., {Piceno}, A., \& {Matthews}, J.~M. 1994, \aj, 108, 678

\bibitem[{{Millan-Gabet} \& {Monnier}(2002)}]{2002mill}
{Millan-Gabet}, R., \& {Monnier}, J.~D. 2002, \apjl, 580, L167

\bibitem[{{Moffat} {et~al.}(2008){Moffat}, {Marchenko}, {Zhilyaev}, {Rowe},
  {Muntean}, {Chen{\'e}}, {Matthews}, {Kuschnig}, {Guenther}, {Rucinski},
  {Sasselov}, {Walker}, \& {Weiss}}]{2008ApJ...679L..45M}
{Moffat}, A.~F.~J., {et~al.} 2008, \apjl, 679, L45

\bibitem[{{O'Keefe} {et~al.}(2009){O'Keefe}, {Gowanlock}, {McConnell}, \&
  {Patton}}]{2009okee}
{O'Keefe}, P.~J., {Gowanlock}, M.~G., {McConnell}, S.~M., \& {Patton}, D. 2009,
  in Astronomical Society of the Pacific Conference Series, Vol. 411,
  Astronomical Society of the Pacific Conference Series, ed. {D.~A.~Bohlender,
  D.~Durand, and P.~Dowler}, 318

\bibitem[{{Paunzen} \& {Reegen}(2008)}]{2008CoAst.153...49P}
{Paunzen}, E., \& {Reegen}, P. 2008, Communications in Asteroseismology, 153,
  49

\bibitem[{Perryman {et~al.}(1997)Perryman, Lindegren, Kovalevsky, Hoeg,
  Bastian, Bernacca, Cr{\'e}z{\'e}, Donati, Grenon, Van~Leeuwen,
  {et~al.}}]{1997perr}
Perryman, M., {et~al.} 1997, Astronomy and Astrophysics, 323, L49

\bibitem[{Press {et~al.}(2001)Press, Vetterling, Teukolsky, \& Flannery}]{nr}
Press, W., Vetterling, W., Teukolsky, S., \& Flannery, B. 2001, Numerical
  recipes in C++: the art of scientific computing (Cambridge University Press
  New York, NY, USA)

\bibitem[{Quinlan(1996)}]{1996quin}
Quinlan, J. 1996, in Proceedings of the National Conference on Artificial
  Intelligence, 725--730

\bibitem[{{Rodr{\'{\i}}guez} {et~al.}(2000){Rodr{\'{\i}}guez},
  {L{\'o}pez-Gonz{\'a}lez}, \& {L{\'o}pez de Coca}}]{2000A&AS..144..469R}
{Rodr{\'{\i}}guez}, E., {L{\'o}pez-Gonz{\'a}lez}, M.~J., \& {L{\'o}pez de
  Coca}, P. 2000, \aaps, 144, 469

\bibitem[{{Sarro} {et~al.}(2009){Sarro}, {Debosscher}, {L{\'o}pez}, \&
  {Aerts}}]{2009A&A...494..739S}
{Sarro}, L.~M., {Debosscher}, J., {L{\'o}pez}, M., \& {Aerts}, C. 2009, \aap,
  494, 739

\bibitem[{{Scargle}(1982)}]{1982scar}
{Scargle}, J.~D. 1982, \apj, 263, 835

\bibitem[{{Sesar} {et~al.}(2007){Sesar}, {Ivezi{\'c}}, {Lupton}, {Juri{\'c}},
  {Gunn}, {Knapp}, {DeLee}, {Smith}, {Miknaitis}, {Lin}, {Tucker}, {Doi},
  {Tanaka}, {Fukugita}, {Holtzman}, {Kent}, {Yanny}, {Schlegel}, {Finkbeiner},
  {Padmanabhan}, {Rockosi}, {Bond}, {Lee}, {Stoughton}, {Jester}, {Harris},
  {Harding}, {Brinkmann}, {Schneider}, {York}, {Richmond}, \& {Vanden
  Berk}}]{2007AJ....134.2236S}
{Sesar}, B., {et~al.} 2007, \aj, 134, 2236

\bibitem[{{Shin} {et~al.}(2009){Shin}, {Sekora}, \&
  {Byun}}]{2009MNRAS.400.1897S}
{Shin}, M., {Sekora}, M., \& {Byun}, Y. 2009, \mnras, 400, 1897

\bibitem[{Silla \& Freitas(2011)}]{2010sill}
Silla, C., \& Freitas, A. 2011, Data Mining and Knowledge Discovery, 22, 31

\bibitem[{{Stankov} \& {Handler}(2005)}]{2005stan}
{Stankov}, A., \& {Handler}, G. 2005, \apjs, 158, 193

\bibitem[{{Stetson}(1996)}]{1996stet}
{Stetson}, P.~B. 1996, \pasp, 108, 851

\bibitem[{{Suchkov} {et~al.}(2005){Suchkov}, {Hanisch}, \& {Margon}}]{2005such}
{Suchkov}, A.~A., {Hanisch}, R.~J., \& {Margon}, B. 2005, \aj, 130, 2439

\bibitem[{Tsoumakas \& Katakis(2007)}]{2007tsou}
Tsoumakas, G., \& Katakis, I. 2007, International Journal of Data Warehousing
  and Mining, 3, 1

\bibitem[{{Udalski} {et~al.}(1999){Udalski}, {Soszynski}, {Szymanski},
  {Kubiak}, {Pietrzynski}, {Wozniak}, \& {Zebrun}}]{1999udal}
{Udalski}, A., {Soszynski}, I., {Szymanski}, M., {Kubiak}, M., {Pietrzynski},
  G., {Wozniak}, P., \& {Zebrun}, K. 1999, \actaa, 49, 1

\bibitem[{Vapnik(2000)}]{2000vapn}
Vapnik, V. 2000, The Nature of Statistical Learning Theory (Springer-Verlag,
  New York)

\bibitem[{Vens {et~al.}(2008)Vens, Struyf, Schietgat, D{\v{z}}eroski, \&
  Blockeel}]{2008vens}
Vens, C., Struyf, J., Schietgat, L., D{\v{z}}eroski, S., \& Blockeel, H. 2008,
  Machine Learning, 73, 185

\bibitem[{{Walkowicz} {et~al.}(2009){Walkowicz}, {Becker}, {Anderson}, {Bloom},
  {Georgiev}, {Grindlay}, {Howell}, {Long}, {Mukadam}, {Prsa}, {Pepper}, {Rau},
  {Sesar}, {Silvestri}, {Smith}, {Stassun}, \& {Szkody}}]{2009astro2010S.307W}
{Walkowicz}, L.~M., {et~al.} 2009, arXiv e-print 0902.3981

\bibitem[{Wasserman(2006)}]{2006wass}
Wasserman, L. 2006, All of nonparametric statistics (Springer-Verlag, New York)

\bibitem[{{Whitelock} {et~al.}(2008){Whitelock}, {Feast}, \& {van
  Leeuwen}}]{2008MNRAS.386..313W}
{Whitelock}, P.~A., {Feast}, M.~W., \& {van Leeuwen}, F. 2008, \mnras, 386, 313

\bibitem[{{Willemsen} \& {Eyer}(2007)}]{2007arXiv0712.2898W}
{Willemsen}, P.~G., \& {Eyer}, L. 2007, arXiv e-print 0712.2898

\bibitem[{{Wo{\'z}niak} {et~al.}(2004){Wo{\'z}niak}, {Williams}, {Vestrand}, \&
  {Gupta}}]{2004AJ....128.2965W}
{Wo{\'z}niak}, P.~R., {Williams}, S.~J., {Vestrand}, W.~T., \& {Gupta}, V.
  2004, \aj, 128, 2965

\bibitem[{Wu {et~al.}(2004)Wu, Lin, \& Weng}]{2004wu}
Wu, T., Lin, C., \& Weng, R. 2004, The Journal of Machine Learning Research, 5,
  975

\bibitem[{{Zechmeister} \& {K{\"u}rster}(2009)}]{zech09}
{Zechmeister}, M., \& {K{\"u}rster}, M. 2009, \aap, 496, 577

\end{thebibliography}

\end{document}